# Enormous Berry-Curvature-Driven Anomalous Hall Effect in Topological Insulator (Bi,Sb)$_2$Te$_3$ on Ferrimagnetic Europium Iron Garnet beyond 400 K

Wei-Jhih Zou[1,#], Meng-Xin Guo[1,#], Jyun-Fong Wong[1,#], Zih-Ping Huang[2,#], Jui-Min Chia[1,#], Wei-Nien Chen[1], Sheng-Xin Wang[1], Keng-Yung Lin[2], Lawrence Boyu Young[2], Yen-Hsun Glen Lin[2], Mohammad Yahyavi[3], Chien-Ting Wu[4], Horng-Tay Jeng[1,5,6], Shang-Fan Lee[5], Tay-Rong Chang[3,6,7*], Minghwei Hong[2*], Jueinai Kwo[1*]

[1]Department of Physics, National Tsing Hua University, Hsinchu 30013, Taiwan

[2]Graduate Institute of Applied Physics and Department of Physics, National Taiwan University, Taipei 10617, Taiwan

[3]Department of Physics, National Cheng Kung University, Tainan 701, Taiwan

[4]Materials Analysis Division, Taiwan Semiconductor Research Institute, National Applied Research Laboratories, Hsinchu 300091, Taiwan

[5]Institute of Physics, Academia Sinica, Taipei 11529, Taiwan

[6]Physics Division, National Center for Theoretical Sciences, National Taiwan University, Taipei 10617, Taiwan

[7]Center for Quantum Frontiers of Research and Technology (QFort), Tainan 701, Taiwan

[#]W.-J. Z., M.-X. G., J.-F. W., Z.-P. H., and J.-M. C. contributed equally to this work.

[*]Address correspondence to J. Kwo, raynien@phys.nthu.edu.tw; M. Hong, mhong@phys.ntu.edu.tw; T.-R. Chang, u32trc00@phys.ncku.edu.tw

ABSTRACT.

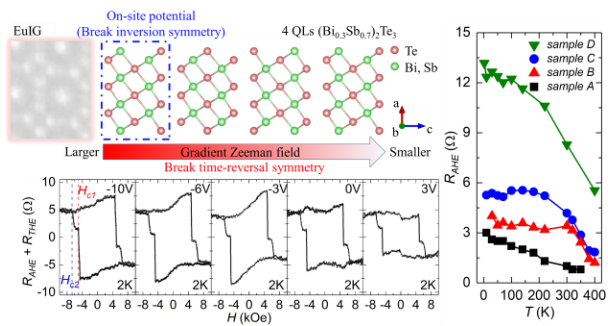

To realize the quantum anomalous Hall effect (QAHE) at elevated temperatures, the approach of magnetic proximity effect (MPE) was adopted to break the time-reversal symmetry in the topological insulator $(Bi_{0.3}Sb_{0.7})_2Te_3$ (BST) based heterostructures with a ferrimagnetic insulator europium iron garnet (EuIG) of perpendicular magnetic anisotropy. Here we demonstrate phenomenally large anomalous Hall resistance ($R_{AHE}$) exceeding 8 Ω ($\rho_{AHE}$ of 3.2 μΩ·cm) at 300 K and sustaining to 400 K in 35 BST/EuIG samples, surpassing the past record of 0.28 Ω ($\rho_{AHE}$ of 0.14 μΩ·cm) at 300 K. The remarkably large $R_{AHE}$ as attributed to an atomically abrupt, Fe-rich interface between BST and EuIG. Importantly, the gate dependence of the AHE loops shows no sign change with varying chemical potential. This observation is supported by our first-principles calculations *via* applying a gradient Zeeman field plus a contact potential on BST. Our calculations further demonstrate that the AHE in this heterostructure is attributed to the intrinsic Berry curvature. Furthermore, for gate-biased 4 nm BST on EuIG, a pronounced topological Hall effect (THE) coexisting with AHE is observed at the negative top-gate voltage up to 15 K. Interface tuning with theoretical calculations has opened up new opportunities to realize topologically distinct phenomena in tailored magnetic TI-based heterostructures.

INTRODUCTION

A three-dimensional topological insulator (TI) is a nontrivial state of matter hosting insulating bulk and conducting surface states (SSs) protected by the time-reversal symmetry (TRS).[1,2] The breaking of TRS at the SSs of a TI leads to the formation of an exchange gap and the emergence of chiral edge states, which give rise to the quantum anomalous Hall effect (QAHE) with the Hall resistance quantized to $h/e^2$ ($h$ is the Planck constant and $e$ is the elementary charge.) when the Fermi level ($E_F$) is tuned within the exchange gap.[3–5] Although the magnetic doping was proven to be effective in breaking the TRS, the reported QAHE observation temperature so far was less than 2 K.[4,6–10] However, the inherent spin disorder due to the random magnetic dopants, as well as the small size of the exchange gap induced by the low doping concentration, may pose an ultimate limit in raising the QAHE temperature.[10,11]

Recently, the magnetic proximity effect (MPE) of TI/ferrimagnetic insulator (FI) heterostructures has been demonstrated to be another route of breaking the TRS.[12–15] When a nonmagnetic TI contacts an FI with the magnetic moment perpendicular to the interface, the magnetic exchange interaction *via* the interface can open an exchange gap at the SSs of the TI. In contrast to magnetic doping, the MPE offers advantages such as fewer crystal defects causing less spin disorder, more uniformly induced interfacial magnetization, and possibly a higher ferromagnetic Curie temperature ($T_c$), thus paving a pathway for realizing the QAHE at significantly higher temperatures. To date, the coupling *via* MPE has been reported in several notable FIs [EuS,[13] Y$_3$Fe$_5$O$_{12}$ (YIG),[14] and Tm$_3$Fe$_5$O$_{12}$ (TmIG)[16–18]] interfaced to TIs with the observation of anomalous Hall effect (AHE). For example, TmIG is a high $T_c$ (560 K) rare-earth iron garnet, and epitaxial TmIG films were grown on (111) oriented Gd$_3$Ga$_5$O$_{12}$ (GGG) by pulsed laser deposition and magnetron sputtering.[19–22] The magnetoelastic anisotropy due to the epitaxial strains in TmIG films on GGG-based substrates leads to the perpendicular magnetic anisotropy (PMA) of TmIG. The exchange coupling between (Bi$_{0.3}$Sb$_{0.7}$)$_2$Te$_3$ (BST) and TmIG has resulted in a significantly enhanced $T_c$ and the robust ferromagnetism in BST with distinctly squarish anomalous Hall hysteresis loops up to 400 K.[16]

Aside from TmIG, another rare-earth iron garnet Eu$_3$Fe$_5$O$_{12}$ (EuIG) with a similar $T_c$ has gained much attention lately because of its stronger PMA field and larger coercive field ($H_c$) reported.[23,24] EuIG thin films by off-axis sputtering recently achieved by Guo *et al.* have shown ultra-smooth surfaces with robust and tunable PMA.[25] The Eu/Fe composition ratio (*Eu/Fe*) was finely adjusted in order to induce different degrees of compressive strain so that the magnetic properties of EuIG films were effectively manipulated, making EuIG films versatile for various spintronic applications.

In contrast to the QAHE, which originates from the Berry curvature (BC) in the momentum space, chiral spin textures from the BC in the real space provide the topological Hall effect (THE).[26] This transport signature of the non-zero spin chirality is induced by spin textures such as skyrmions stabilized by the Dzyaloshinskii–Moriya interaction (DMI).[26] The DMI at the interface of TI/FI is due to the spin-orbit coupling (SOC) in TI and the broken inversion symmetry of the bilayer. Moreover, the spin-momentum locking of topological SSs makes the interfacial DMI significantly stronger than the heavy-metal-based system.[27] The stronger DMI strength is beneficial to form the small-sized skyrmions, thereby increasing the memory capacity per unit volume in skyrmion-based storage devices.[28]

In this work, we report the attainment of unprecedentedly large anomalous Hall resistance ($R_{AHE}$) in BST/EuIG heterostructures to high temperatures of at least 400 K. By implementing a crucial step of high-temperature annealing to the EuIG films prior to the BST growth, we succeeded in tailoring the BST/EuIG interfaces to be atomically abrupt and considerably Fe-rich. With over 35 samples fabricated, we discovered an enormous enhancement of the $R_{AHE}$ and the $\rho_{AHE}$ exceeding twenty times higher than the previous record at room temperature.[16] Furthermore, the top-gate electrical-field effect performed on BST/EuIG showed no sign change of AHE loops with the chemical potential, consistent with our comprehensive theoretical investigation of the AHE using first-principles and the linear-response method. We identified that the AHE was indeed realized in BST because of its intrinsic Berry curvature, an artificially applied gradient Zeeman field, and an on-site potential, which simulated the MPE and the interface effect in our BST/EuIG heterostructures. Crucially, our calculations further suggest that the nearly quantized AH conductivity (AHC) within the insulating gap begins to round off as the temperature increases and makes a prominent peak of AHC. In addition, a distinct signature of THE was observed primarily at the negative top-gate voltage ($V_{gate}$) up to 15 K. Above discoveries will not only impact the study of novel topological phenomena, including THE and topological magnetoelectric effect (TME), *etc.*, but also have substantial implications for developing dissipation-less spintronic devices in the future.

RESULTS AND DISCUSSION

**Materials growth and characterization of EuIG films and BST/EuIG heterostructures.** Epitaxial EuIG films with PMA were obtained on GGG substrates because of the positive magnetostriction constant and the in-plane compressive strain of EuIG (bulk lattice constants $a_{EuIG}$ = 12.49 Å and $a_{GGG}$ = 12.38 Å).[29,30] By tuning the substrate-to-target distance in the off-axis sputtering configuration, EuIG films with increasing *Eu/Fe* from 0.477, 0.529, to 0.577 were grown. Squarish magnetic hysteresis loops (M-H loops) under out-of-plane magnetic fields in Figure 1a revealed the robust PMA of the EuIG films. Excessive $Fe^{3+}$ in the EuIG films could replace $Eu^{3+}$ in the dodecahedron sites, resulting in a stronger antiferromagnetic coupling in EuIG and a reduced saturation magnetization ($M_s$). A comprehensive account of the sputtering growth, structural and magnetic properties of the EuIG films is reported elsewhere.[25] Figure S1 in Supporting Information shows the atomic force microscopy (AFM) images of as-deposited and annealed EuIG films with smooth surface morphology and small root-mean-square roughness ($R_q$) ~0.12 nm, essential to form an abrupt BST/EuIG interface.

BST (001) thin films were grown on EuIG (001) by an approach following the Se-buffered low-temperature (SBLT) growth method,[17] where Te was substituted for Se to grow Te-based TIs. In addition, a high-temperature annealing process was implemented prior to the BST growth to improve the starting EuIG surface conditions. A EuIG film annealed at 700°C for 30 minutes has led to a bright and streaky reflection high-energy electron diffraction (RHEED) pattern in Figure 1b, indicating a flat and well-ordered surface. Besides, the carbon contamination on the EuIG surface due to air exposure was significantly reduced to a minor level as diagnosed by X-ray photoelectron spectroscopy (XPS) in Figure S2a in Supporting Information. BST thin films with a typical thickness of 7 nm were obtained, evidenced by the sharp and streaky RHEED pattern in Figure 1c. Additional BST/EuIG samples were grown with various annealing temperatures ($T_a$) from 450 °C to 750 °C on EuIG. The corresponding AFM images of the samples are shown in Figure S3 in Supporting Information, and the smallest $R_q$ of BST/EuIG was obtained for $T_a$ of 650 °C.

Figure 1d shows the X-ray diffraction (XRD) scan along the surface normal of BST/EuIG/GGG with all the diffraction peaks indexed, and the $c$ lattice constants of EuIG and GGG were determined to be 12.548 Å and 12.382 Å, respectively. The former value is larger than $a_{EuIG}$, suggesting an out-of-plane tensile strain in EuIG. The reciprocal lattice map (not shown) of off-normal reflections evidenced that EuIG is fully strained on GGG, and the EuIG lattice is thus tetragonally distorted with an elongated $c$-axis. Pronounced Pendellösung fringes in Figure 1d manifest the sharp interface between EuIG and GGG, and the long-range order of EuIG along the $c$-axis. Figure 1e shows the XRD azimuthal φ scans crossing GGG and EuIG off-normal {204} reflections with peaks separated by 90°, indicating the four-fold symmetry of both lattices. Moreover, the coincidence of GGG {204} and EuIG {204} reflections in the φ scans revealed the in-plane alignment between the two lattices. On the other hand, the φ scan crossing BST {105} (with a three-fold symmetry) shows diffraction peaks separated by 30° instead of 120°. This finding reveals that BST consists of four rotational domains and each of them has its (100) planes aligned with one of the four EuIG {100} planes.

The spherical aberration-corrected scanning transmission electron microscopy (Cs-STEM) high-angle annular dark-field (HAADF) image in Figure 1f reveals the atomic arrangement of the sharp BST/EuIG interface with a van der Waals (vdW) gap observed. The BST stemmed from the initial Te layer on the EuIG with quintuple layer (QL) lamellae also separated by vdW gaps. The atomic arrangement of Te–Sb(Bi)–Te–Sb(Bi)–Te in a QL was confirmed by the intensity profile in Figure 1g. STEM electron energy loss spectroscopy (STEM-EELS) spectra of EuIG probed at positions 1 to 4 denoted in Figure 1f are shown in Figure 1i,j. The intensities of Fe $L_{3,2}$ and Eu $M_{5,4}$ were significantly decreased when the probe moved from 1 to 2 (or 3 to 4), consistent with the darker area near the top of EuIG in the Cs-STEM HAADF image. Importantly, the intensity of Eu $M_5$ was reduced more than that of Fe $L_3$, leading to the intensity ratio of $I_{Fe}/I_{Eu}$ increasing from 1.56 to 2.8 (or 1.8 to 4.5), summarized in Figure 1h. These results indicate lower Eu and Fe densities near the top of EuIG, and there are more Eu vacancies. Substantially Fe-rich (over Eu) layer at the top of EuIG ~0.6 nm from the interface could be due to a preferential loss of $EuO_x$ during the high-temperature annealing. Higher-$T_a$ annealed EuIG films exhibiting Fe-richer surfaces were also found by XPS, as illustrated in Figure S2b in Supporting Information.

**Tunability of enormous AHE at room temperature by varying growth parameters of *Eu/Fe* and $T_a$.** Temperature-dependent Hall effect measurements were conducted on a series of BST/EuIG samples. In general, the EuIG and BST thicknesses were kept at 23 nm and 7 nm, respectively. The Hall traces of *sample A* (*Eu/Fe* = 0.529, $T_a$ = 450 °C) from 10 K to 350 K are shown in Figure 2a. As the temperature increased, the slope of the ordinary Hall effect (OHE) turned from positive to negative, in which the dominant carriers gradually changed from holes to electrons. The nonlinearity of the Hall traces was the evidence of ambipolar transport, indicating that the $E_F$ of BST was very close to the exchange gap and far away from the bulk bands. Moreover, pronounced hysteresis loops observed in all the Hall traces illustrated the existence of AHE up to 350 K. The Hall resistances ($R_{HE}$) are only contributed from the BST film as EuIG is an insulating layer, and the AHE signals shown in Figure 2b were extracted by subtracting a linear OHE background between ±10 kOe. An enormous $R_{AHE}$ was attained ~0.80 Ω ($\rho_{AHE}$ ~0.56 μΩ·cm) at 350 K. In contrast, the previous $R_{AHE}$ record in BST/TmIG was ~0.18 Ω ($\rho_{AHE}$ ~0.09 μΩ·cm) at the same temperature.[16] Note that the sign of AHE in BST/EuIG remained negative despite that the conduction carriers were altered from holes to electrons with the increasing temperature. This finding is similar to the published work on BST/TmIG,[16] and will be addressed further in the gate-dependent results in Figure 4. The squarish AHE loops of BST/EuIG originated from the MPE-induced ferromagnetism in BST with a gapped bottom SS due to the strong PMA of EuIG. The observation of AHE up to 350 K was benefited from the robust PMA of EuIG with a high $T_c > 400$ K.

The MPE-induced ferromagnetism in BST was realized through the exchange coupling between BST and EuIG; therefore, the interface played a critical role in leading the large AHE above room temperature. To achieve even stronger exchange coupling and AHE by tailoring the BST/EuIG interface, EuIG films were heated up to the elevated $T_a$ prior to the BST growth. Moreover, the effect of *Eu/Fe* on the AHE strength was investigated at the same time. Figure 2c,d show the $T_a$-dependent $R_{AHE}$ and $H_c$ in BST/EuIG with different *Eu/Fe* of 0.477, 0.529, and 0.577.

Notably, unprecedentedly large $R_{AHE}$ values of ~3.41 Ω and ~4.17 Ω were obtained at 300 K in *sample B* ($Eu/Fe = 0.529$, $T_a = 650$ °C) and *sample C* ($Eu/Fe = 0.477$, $T_a = 700$ °C), respectively.

The $T_a$-dependent $R_{AHE}$ and $H_c$ in Figure 2c,d exhibit a similar trend for different $Eu/Fe$, showing an increase for $T_a$ from 450 °C to 700 °C, and then a drop at 750 °C. In addition, BST grown on Fe-richer EuIG (blue dots) in general shows larger $R_{AHE}$ and $H_c$, suggesting that $R_{AHE}$ values could be correlated to the net magnetic moments of $Fe^{3+}$ in EuIG. One published work on $Zn_{1-x}Cr_xTe$/BST/$Zn_{1-x}Cr_xTe$ has reported the MPE-induced QAHE by coupling between 3*d*-orbitals of transition metals in $Zn_{1-x}Cr_xTe$ and 5*p*-orbitals of Te atoms in BST.[31] Hence, the pronounced AHE in BST/EuIG could be attributed to the Fe-rich top layer in the annealed EuIG detected by STEM-EELS, thus favoring the coupling between 3*d*-orbitals of Fe and 5*p*-orbitals of Te.

Figure 2e displays the AHE loops of BST/EuIG measured at 300 K with $Eu/Fe = 0.477$ and increasing $T_a$ from 450 °C to 750 °C, and all of them show sizable $R_{AHE}$. The AHE was enhanced significantly in BST/EuIG with $T_a$ increased from 450 °C to 700 °C, which could be attributed to the cleaner EuIG surface with the significantly reduced carbon contamination, as well as the Fe-richer EuIG surface evidenced by the XPS data in the inset of Figure S2b in Supporting Information. However, a sudden drop of $R_{AHE}$ occurred for BST/EuIG with $T_a = 750$ °C, and this might be caused by the cracks of the EuIG surface shown in Figure S3f in Supporting Information due to the overheating. Overall, the tunability of AHE strength in BST/EuIG by varying the $T_a$ and the $Eu/Fe$ of EuIG have been achieved, potential for applications in room-temperature spintronics.

**Temperature dependence of anomalous Hall effect.** To maximize the $R_{AHE}$, Fe-richer EuIG with $Eu/Fe$ = 0.477 was chosen to be the FI layer for stronger coupling between BST and EuIG. Figure 3a manifests the AHE loops of *sample C* measured from 300 K to 400 K, and the $R_{AHE}$ retained a large value of ~1.84 Ω at 400 K. Hall effect data above 400 K were not acquired due to the instrument limitation. The giant $R_{AHE}$ values not only stand for the robustness of MPE up to 400 K but also indicate an abrupt interface between BST and EuIG with intimate contact. The squareness (*SQR*) of an AHE loop is defined as $SQR = \frac{R_r}{R_s}$ here, where $R_r$ and $R_s$ are the $R_{AHE}$ at the zero field and the switching field, respectively. The *SQR* values of all the BST/EuIG samples are approximately equal to 1.0.

The $R_{AHE}$ values of *samples A*, *B*, *C*, and *D* from 10 K to 400 K are shown in Figure 3b. The highest records from previous publications over a similar temperature range are also included for comparison.[16] The $R_{AHE}$ values of BST/EuIG in this work surpass the previous records over one order of magnitude above 300 K. The stronger PMA in EuIG than that in TmIG could induce an enhanced spontaneous magnetization in the topological SSs of BST. The attainment of larger $R_{AHE}$ values in BST/EuIG than those in BST/TmIG is likely caused by the outstanding interfacial quality and the stronger PMA of EuIG. The high-temperature AHE arising from the MPE in BST/EuIG and BST/TmIG implies that the $T_c$ is above 400 K, notably superior to those of Cr- and V-doped BST reported.[4,8,10,32] Hence, inducing a magnetic order in topological materials *via* MPE holds a good prospect to realize QAHE at higher temperatures than that *via* magnetic doping. (See discussions in Section S11, Supporting Information) Importantly, the BST/EuIG heterostructures with sharp interfaces can be utilized as a platform to investigate novel topological quantum phenomena. For example, a BST (4 nm)/EuIG (*sample D*, *Eu/Fe* = 0.477, $T_a$ = 700 °C) was fabricated where the two surfaces are hybridized. A striking $R_{AHE}$ value doubles to ~8 Ω at 300 K and monotonically increases to ~13 Ω at 2 K; other characteristic features associated with THE emerge, as shown later in Figure 4b and 6a.[27,33,34] Moreover, we discovered an interesting trend that the $α$ values extracted from the Hikami–Larkin–Nagaoka (HLN) equation gradually reduce to zero while the $R_{AHE}$ values are larger, as shown in Figure S4b in Supporting Information. The positive correlation between the $R_{AHE}$ and the $α$ values suggests that the strong PMA of EuIG magnetizes both the bottom and the top SSs of BST *via* decreasing the BST thickness in *sample D*.

The temperature-dependent $H_c$ values of the AHE loops of *samples A*, *B*, *C*, and *D* are shown in Figure 3c. The robust PMA sustained at an elevated temperature with $H_c$ ~0.3 kOe for *sample C* at 400 K. The $H_c$ values in all these four samples increased significantly with decreasing temperatures. Note that the temperature-dependent $H_c$ of M-H loops for EuIG is expected to follow the equation that describes the thermal activation of domain walls, $H_c = H_0 \left(1 - (\frac{T}{T_B})^{0.5}\right)$, where $H_0$ is the coercive field at 0 K and $T_B$ is the blocking temperature.[35] In Figure 3c, the temperature-dependent $H_c$ of the AHE follows closely with this equation, suggesting that the AHE of BST/EuIG is caused by interface effects owing to EuIG, such as MPE or spin Hall effect (SHE). Mechanisms responsible for MPE- and SHE-induced AHE are related to exchange coupling and SOC, respectively.[36] To discern the correct mechanism underlying the AHE in BST/EuIG, we investigated the dependence of the magneto-resistance (MR) on the magnetic field in out-of-plane and in-plane directions as detailed in Section S5, Supporting Information. Markedly different trends were found between BST/EuIG and Pt/EuIG, where AHE in the latter resulted from SHE. Our analysis thus verified that AHE in BST/EuIG was induced exclusively by MPE.

**Gate-dependent anomalous Hall effect.** To correlate the spin transport property with the electronic structure, *i.e.*, the BC near the $E_F$ in the momentum space, the sign of the Berry-phase associated AHE component may be exploited by employing a top-gate field effect to fine-tune the chemical potential of 4 nm thick BST on EuIG (*sample E*) in the vicinity of the charge neutrality point (CNP). Here the electrical field extends from the top gate toward the BST layer over the entire 4 nm thickness, where the top and the bottom SSs are coupled. Detailed fabrication and measurements of the top-gate devices are described in Section S6, Supporting Information.

Figure 4a shows the longitudinal resistance ($R_{xx}$) with respect to the applied $V_{gate}$. Maximum $R_{xx}$ ~8.5 kΩ occurred at the CNP ($V_{gate} = V_{CNP} = -1.4$ V) and $R_{xx}$ decreased in both hole-doped ($V_{gate} < V_{CNP}$) and electron-doped ($V_{gate} > V_{CNP}$) regions. The small $V_{CNP}$ value indicated that the $E_F$ of the ungated BST was very close to the exchange gap.

Gate-dependent AHE loops are shown in Figure 4b, and two $H_c$'s ($H_{c1}$ and $H_{c2}$) appear in all AHE loops. Here $H_{c1}$ and $H_{c2}$ are attributed to the coercive fields of the hybridized bottom and top surfaces of BST, respectively. The $H_{c1}$ and $H_{c2}$ in all the AHE loops remained constant (~4.7 kOe and ~6.5 kOe) over the applied $V_{gate}$ range, independent of $V_{gate}$ or the $E_F$ of BST, as arising from the thermal activation of the domain walls in EuIG. Since the bottom surface of BST is in contact with EuIG, the magnetization of the bottom surface is expected to be bigger than the top surface, which causes the large jump at $H_{c1}$. Moreover, the surface anisotropy of the bottom surface does not manifest itself, and the induced bottom surface magnetization switches with the bulk EuIG. In contrast, the strong surface anisotropy of the top surface, due to its two-dimensional nature, thus switches at higher fields ($H_{c2} > H_{c1}$).

$R_{AHE}$ values of *sample E* decrease from ~5 Ω to ~3 Ω ($\rho_{AHE}$ ~2.0 μΩ·cm to ~1.2 μΩ·cm) when the $E_F$ crosses from the hole-doped region to the electron-doped region as shown in Figure 4c. The $R_{AHE}$ and $R_{xx}$ exhibit a similar behavior that both decrease monotonically into the electron-doped region. The correlation between the $R_{AHE}$ and $R_{xx}$ was plotted in Figure 4d, showing a power-law dependence with an exponent of ~2. To put it another way, the anomalous Hall conductivity ($\sigma_{AH}$) was nearly independent of the longitudinal conductivity ($\sigma_{xx}$), as shown in Figure 4e. The $\sigma_{AH}$ can be theoretically separated into two probable contributions with different transport lifetime ($\tau$) or $\sigma_{xx}$ dependence. One is proportional to $\tau$, which can only be attributed to the skew-scattering mechanism; the other is proportional to $\tau^0$, which is attributed to the non-zero Berry phase or the side jump mechanism.[37,38] Therefore, the relation of $R_{AHE} \sim R_{xx}^2$ or $\sigma_{AH} \sim \sigma_{xx}^0$ indicated that $\sigma_{AH}$ was unrelated to $\tau$, thus ruling out the skew-scattering mechanism in BST/EuIG. Notably, the sign of $R_{AHE}$ remained negative even though the slope of OHE changed sign in *sample E*, according to the gate-dependent Hall data shown in Figure S7 in Supporting Information. This result is consistent with the temperature-dependent Hall data of *sample A* in Figure 2a. The unchanged sign of the AHE in BST/EuIG heterostructures results from the broken TRS and spatial inversion symmetry, which will be discussed later in Session 2.5.

**Density functional theory calculations on the AHE in 4 QLs BST.** To deepen our understanding the nature of unusually large AHE in BST/EuIG, we calculated the AHE of 4 QLs BST slab model based on the gauge-invariant BC in the momentum space *via* the density functional theory (DFT) calculations associated with the Kubo-formula approach.[39,40] EuIG is a magnetic insulator which will provide an exchange field into the covered BST thin-film to break the TRS, and the interaction between EuIG and BST will modify the interface potential of the hybrid system to break the spatial inversion symmetry in BST. In order to match the experimental settings as much as possible, we artificially applied a gradient Zeeman and on-site energy in the 4 QLs slab model to simulate the decayed exchange field from EuIG into BST and the interface potential of the BST/EuIG interface, respectively. The calculation model with detailed parameters is illustrated in Figure 5a. Figure 5b shows the first-principles-calculated momentum- and energy-resolved BC and the corresponding AHC of 4 QLs BST. The blue and red color on the band structure denotes the positive and negative value of BC, respectively. The largest AHC appears at the gapped Dirac point (DP) because of the band inverted gap. This large and nearly quantized AHC indicates that this inverted Dirac gap carries a non-zero Chern number ($C$), $C = 1$. It is worthy to note that the BC is not selectively strong near the DP but distributes in the wide range of the energy band. Thus, the sign of AHC remains the same and extends from -0.1 eV to 0.1 eV, which is consistent with our gate-dependent measurements. The AHC calculated from the Kubo formula is responsible for the intrinsic AHE, which further supports that the MPE-induced AHE in BST/EuIG is less likely caused by other disorder effects. The temperature-dependent calculations in Figure 5c reveal that the peak of AHC around the DP will be strongly suppressed with increasing temperature but persist even above room temperature, which is attributed to the topological properties of the gap near the Γ point. In addition, the magnitude of the calculated $\sigma_{xy}$ is larger than the experimental $\sigma_{AH}$ values by roughly one order of magnitude. Presumably it is caused by the non-ideal interface between BST and EuIG inducing a somewhat decreased Zeeman field strength and a weaker magnetization in BST.

**Gate- and temperature-dependent topological Hall effect.** AHE loops of *sample E* in Figure 4b reveal humps near the $H_{c1}$ that gained strength with $V_{gate}$ varying from -10 V to +3 V, identifiable to be the THE feature, a transport signature of non-zero spin chirality described by Dzyaloshinskii–Moriya interaction (DMI).[26] Moreover, the second hump feature remains nearly unchanged with varying $V_{gate}$, along with other small humps revealed between $H_{c1}$ and $H_{c2}$. As discussed in Section S10, Supporting Information, after subtracting the AHE component by using the formula, the extracted THE resistance ($R_{THE}$) is ~4.5 Ω ($\rho_{THE}$ ~1.8 μΩ·cm) as shown in Figure 6a. The gate-dependent $R_{THE}$ at 2 K is displayed in Figure 6b. The THE feature is more pronounced in the hole-doped region, and $R_{THE}$ has a maximum value ~4.5 Ω at $V_{gate}$ of -3 V.

Recently Jiang *et al.* reported in gated Cr-BST/BST/Cr-BST sandwiched heterostructures the appearance of THE in the QAH insulating regime resulting from the gate-induced DMI during the magnetization reversal process.[34] Their calculations found a large spin susceptibility that emerges from the bulk valence band to enhance the DMI significantly.[34] On the same token, our observation of gate-induced THE in the AHE regime at negative $V_{gate}$ may also be attributed to the same reason when the $E_F$ of BST is located in the bulk valence band.

Figure 6c shows the gate-dependent $R_{AHE}$ and $R_{THE}$ in temperatures varying from 2 K to 15 K. In all temperatures, $R_{THE}$ has a maximum value for $V_{gate}$ at -3 V, and the hump feature becomes smaller or even disappears as the temperature increases. The extracted $R_{THE}$ values from Figure 6c are shown in Figure 6d at the selected $V_{gate}$. The $R_{THE}$ in all $V_{gate}$ decreases with increasing temperatures, and when the $V_{gate}$ is at -8 V and -10 V, THE feature disappears at 15 K and 10 K, respectively.

The temperature dependence of the THE features is described by reduced DMI strength at a higher temperature. When the thermal fluctuation is larger than the energy scale of DMI, the chirality of the magnetic domain walls would be destroyed, which causes the vanished $R_{THE}$ at the high temperature. Overall, the temperature dependence of THE is generally consistent with that of AHE as expected. Note that the magnitudes of our $\rho_{AHE}$ and $\rho_{THE}$ are ~2.0 and ~1.8 μΩ·cm, respectively at 2 K, about eight to eleven times larger than those of $Bi_2Se_3$/$BaFe_{12}O_{19}$.[27] The novel interplay between the chiral edge states and the chiral spin textures in magnetic TI heterostructures calls for further in-depth studies now underway.

CONCLUSIONS

We have demonstrated BST/EuIG heterostructures with outstanding materials characteristics, magneto-transport features, and theoretical calculations. Notably, the sign of $R_{AHE}$ remains negative with varying chemical potential, and the enormous magnitude of $\rho_{AHE}$ exceeds twenty times higher than the previous record at room temperature, sustaining to 400 K. These striking results are well interpreted by our DFT calculations based on an artificially applied Zeeman field plus a contact potential on BST. Therefore, our ability to produce these tailored magnetic TI heterostructures in conjunction with the theoretical guidance for interface tunings has opened up exciting opportunities to realize topologically distinct phenomena such as THE and topological magnetoelectric effect (TME) in axion insulators,[41] *etc*. The coexistence of AHE and THE may provide an excellent opportunity to investigate the correlation between chiral edge states and chiral spin textures. It will lead to TI-based dissipation-less and low-power spintronics in the future, realizing these TI materials for practical applications.

EXPERIMENTAL METHODS

**Materials growth.** Ferrimagnetic insulator EuIG thin films were grown on GGG substrates in (001) orientation using the off-axis magnetron sputtering technique.[25] The EuIG samples were transferred to a standard molecular beam epitaxy (MBE) system with a base pressure of $4 \times 10^{-10}$

Torr, and annealed at temperatures varying from 450 °C to 750 °C. After the EuIG films were cooled to room temperature, the BST film growth commenced by evaporation from high purity (99.9999%) Bi, Sb, and Te sources. The crystallinity of the BST thin film surface was monitored by RHEED patterns during the growth, and the Bi:Sb composition ratio was kept at 3:7 to control the $E_F$ close to the exchange gap with a growth rate of 0.36 nm/min.

**Characterization.** The surface morphologies of the annealed EuIG and BST/EuIG samples were examined by AFM using the non-contact mode. The epitaxial relationship between BST, EuIG, and GGG was studied using synchrotron-radiation XRD at BL17B beamline ($\lambda = 1.5498$ Å) of Taiwan Light Source, Hsinchu, Taiwan. The BST/EuIG interface was characterized by Cs-STEM using a HAADF detector. The experiments were performed on an aberration-corrected (a 0.9 Å probe size) JEOL 2100F, operated at an accelerating voltage of 200 kV. The STEM samples were prepared by using mechanical polishing and focused ion beam. The atomic models in Figure 1f were drawn with the VESTA software. Notice that the excessive oxygen positions in the structural model on the STEM image are not shown for simplicity.

**Electrical measurements.** The BST/EuIG samples were patterned into Hall bar devices (650×50 μm$^2$) by photolithography and reactive ion etching. Four-terminal Hall measurements from 10 K to 400 K were conducted in a Quantum Design Physical Property Measurement System (PPMS) using a 30 μA direct current source. The top-gate sample fabrication and measurements are described in Section S6, Supporting Information. In Hall measurements, $R_{xx}$ often mixes with $R_{HE}$ because of the misalignment of contact electrodes. To remove this effect, the "anti-symmetrization method" was carried out for all Hall traces $R_{HE}(H) = \frac{R_{HE}^{raw}(H) - R_{HE}^{raw}(-H)}{2}$.

**First-principles calculations in BST.** First-principles electronic property calculations of $Bi_2Te_3$ ($Sb_2Te_3$) were performed using the projector augmented wave (PAW) potentials within the standard DFT framework as implemented in the Vienna *ab initio* Simulation Package (VASP).[42–44] In the calculations of electronic structures, the SOC was included self-consistently with a MonkhorstPack k-point grid of size 13 × 13 × 5. To systematically calculate the bulk and surface electronic structures, and AHC in few QLs of $(Bi,Sb)_2Te_3$, we constructed a tight-binding Hamiltonian for both $Bi_2Te_3$ and $Sb_2Te_3$ by projecting onto the Wannier orbitals,[45–47] which used the VASP2WANNIER90 interface.[48] For generating the Wannier functions of a real space tight-binding model of $Bi_2Te_3$ and $Sb_2Te_3$, we used Bi (*p*-orbitals), Sb (*p*-orbitals), and Te (*p*-orbitals) without performing the procedure for maximally localized Wannier functions. The electronic structure of the $(Bi,Sb)_2Te_3$ with Bi:Sb = 3:7 was calculated by linear interpolation of tight-binding model matrix elements of $Bi_2Te_3$ and $Sb_2Te_3$. This approach was successfully applied to investigate the evolution of band topology in $BiTlSe_{1-x}S_x$ TIs and $Mo_xW_{1-x}Te_2$ Weyl semimetals. In addition, when the chemical potential is away from the exchange gap and crosses the bulk bands, it is not clear whether the intrinsic AHE is still dominant in this system. The AHC calculated in this work is in a range of chemical potential from -0.1 eV to 0.1 eV, near or located in the exchange gap.

FIGURES.

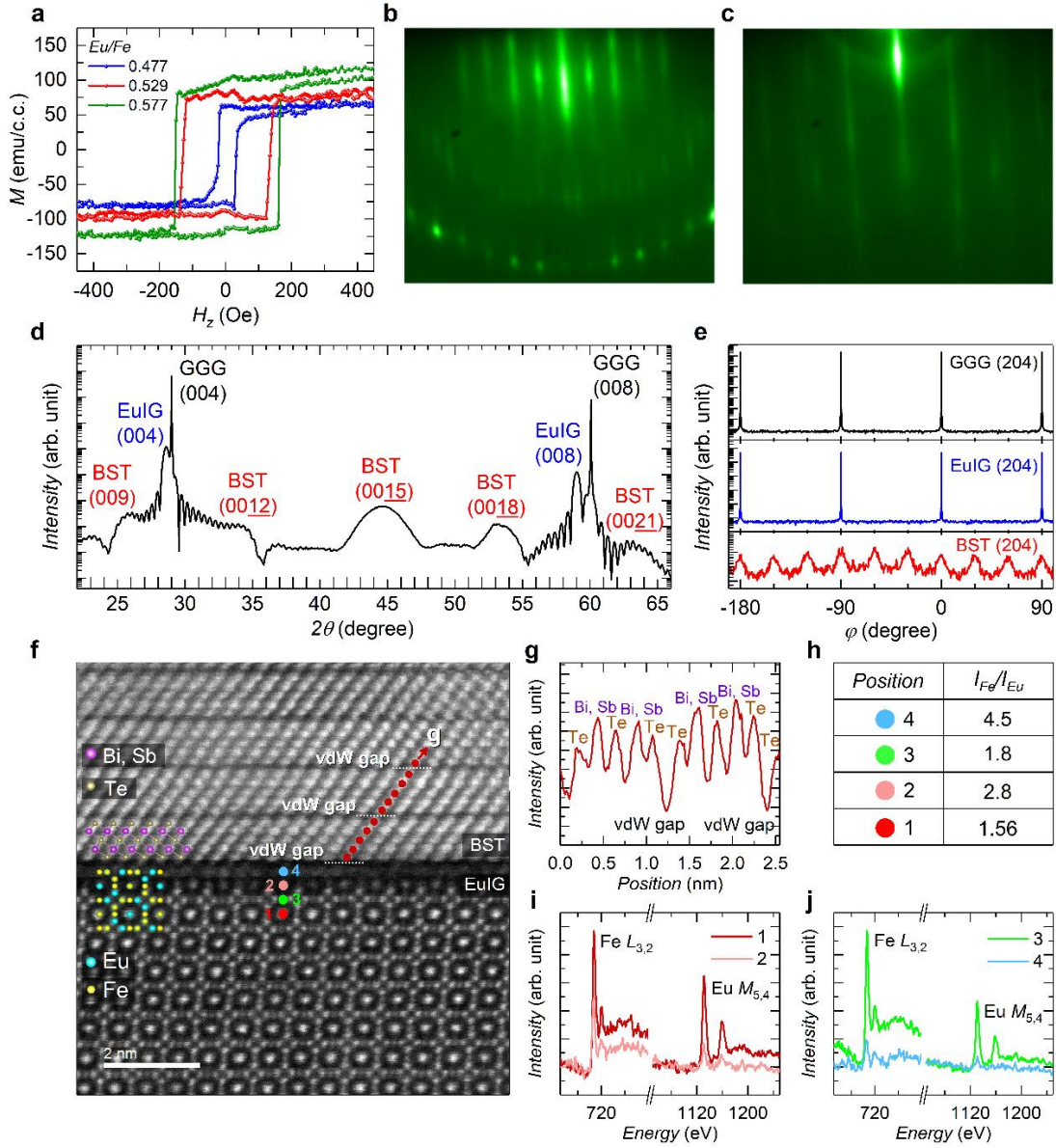

**Figure 1.** Magnetic and structural analysis of BST/EuIG. (a) M-H loops of EuIG films of different *Eu/Fe* with out-of-plane applied magnetic fields. (b) and (c) are the RHEED patterns of EuIG (001) surface annealed at 700 °C for 30 minutes along the [110] axis, and 7 nm BST (001) surface along the [100] axis, respectively. (d) XRD scan along the surface normal and (e) azimuthal φ scans

crossing GGG {204}, EuIG {204}, and BST {105} reflections of BST/EuIG/GGG(001) using X-rays of 1.5498 Å wavelength. Four-domain BST was grown on single-domain EuIG with an orientation relationship of BST{100}//EuIG{100}//GGG{100}. (f) Cs-STEM HAADF image of BST/EuIG. The QLs of BST are separated by vdW gaps denoted, and the atomic arrangement in a QL (along the red arrow) is Te–Sb(Bi)–Te–Sb(Bi)–Te, which is confirmed by the intensity profile shown in (g). (i) and (j) are the position-dependent STEM-EELS spectra of EuIG near Fe $L_{3,2}$ and Eu $M_{5,4}$ edges probed at positions 1 to 4 denoted in (f), and the intensity ratios are summarized in (h).

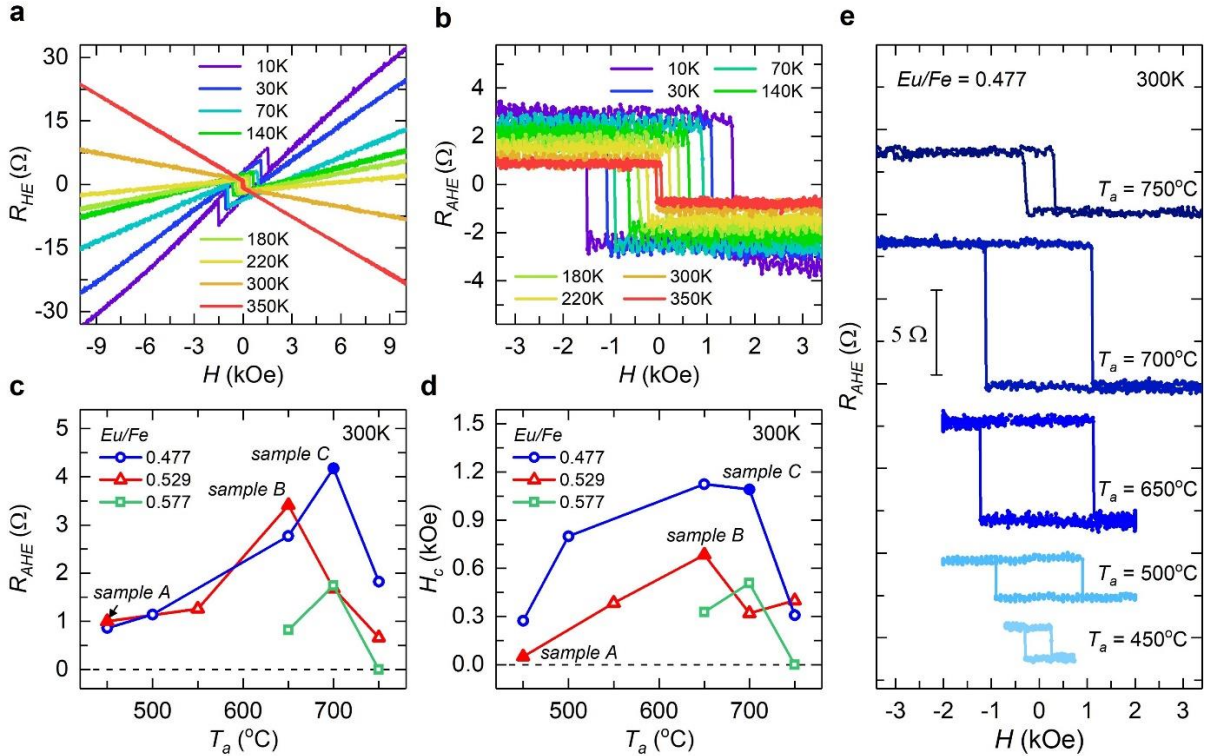

**Figure 2.** AHE loops varied with temperatures and growth parameters of BST/EuIG. (a) $R_{HE}$ vs magnetic field for *sample A*. The Hall loops are visible in the low field region. (b) AHE loops of *sample A* at measurement temperatures from 10 K to 350 K. (c) $R_{AHE}$ (300 K) and (d) $H_c$ (300 K) for samples prepared with various $T_a$ and *Eu/Fe*, the solid symbols refer to *samples A*, *B*, and *C*. (e) AHE loops (300 K) of samples with *Eu/Fe* = 0.477 and various $T_a$.

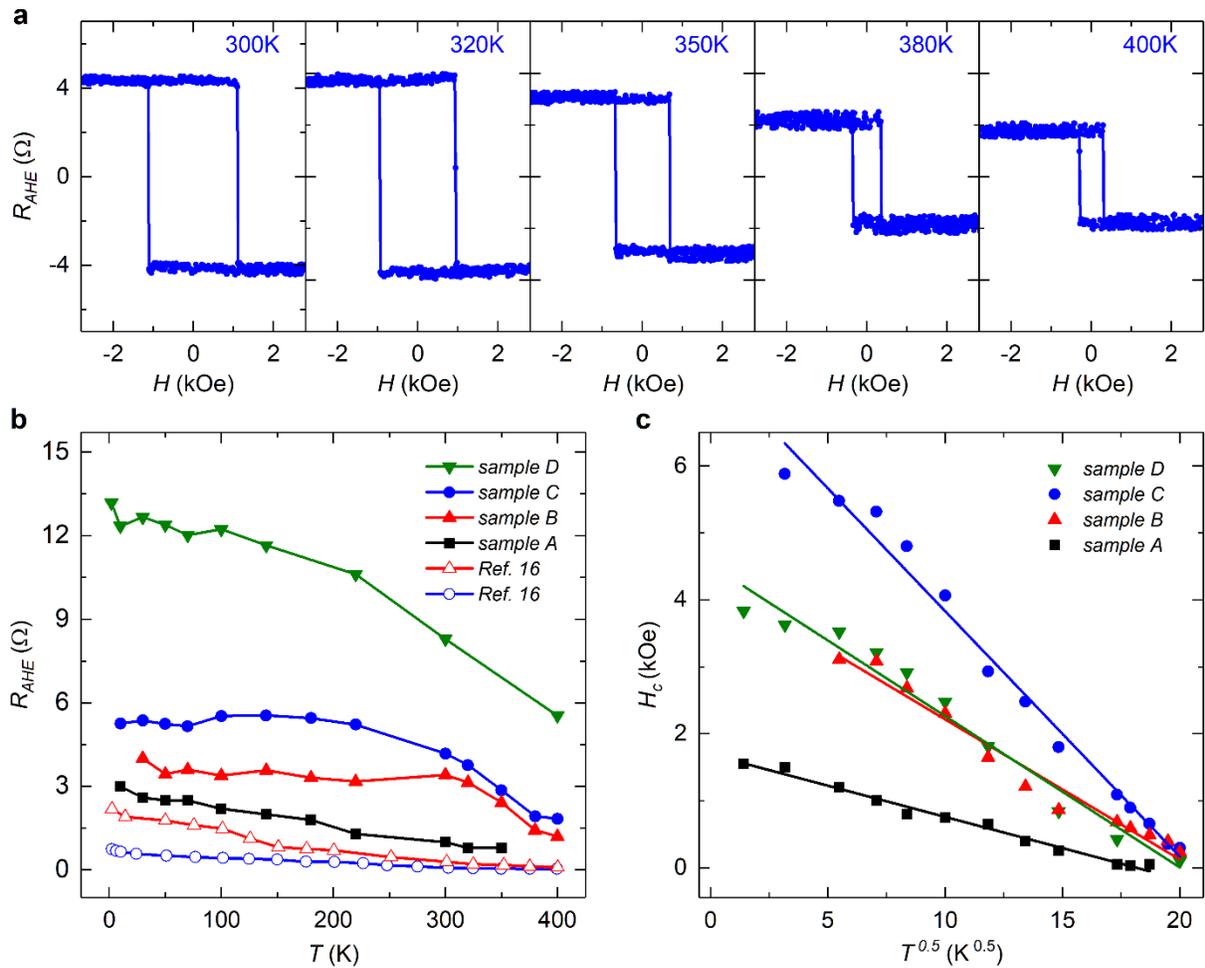

**Figure 3.** Temperature-dependent AHE properties of BST/EuIG. (a) AHE loops of *sample C*. (b) $R_{AHE}$ of four *samples A*, *B*, *C*, and *D* and the other two datasets from C. Tang *et al.*[16] Note that at 300 K, $\rho_{AHE}$ of *sample D* is ~3.2 µΩ·cm, and that of the highest record from previous publications is ~0.14 µΩ·cm.[16] (c) $H_c$ of four samples in this work. The straight lines plotted in (c) are the linear fits of the data points.

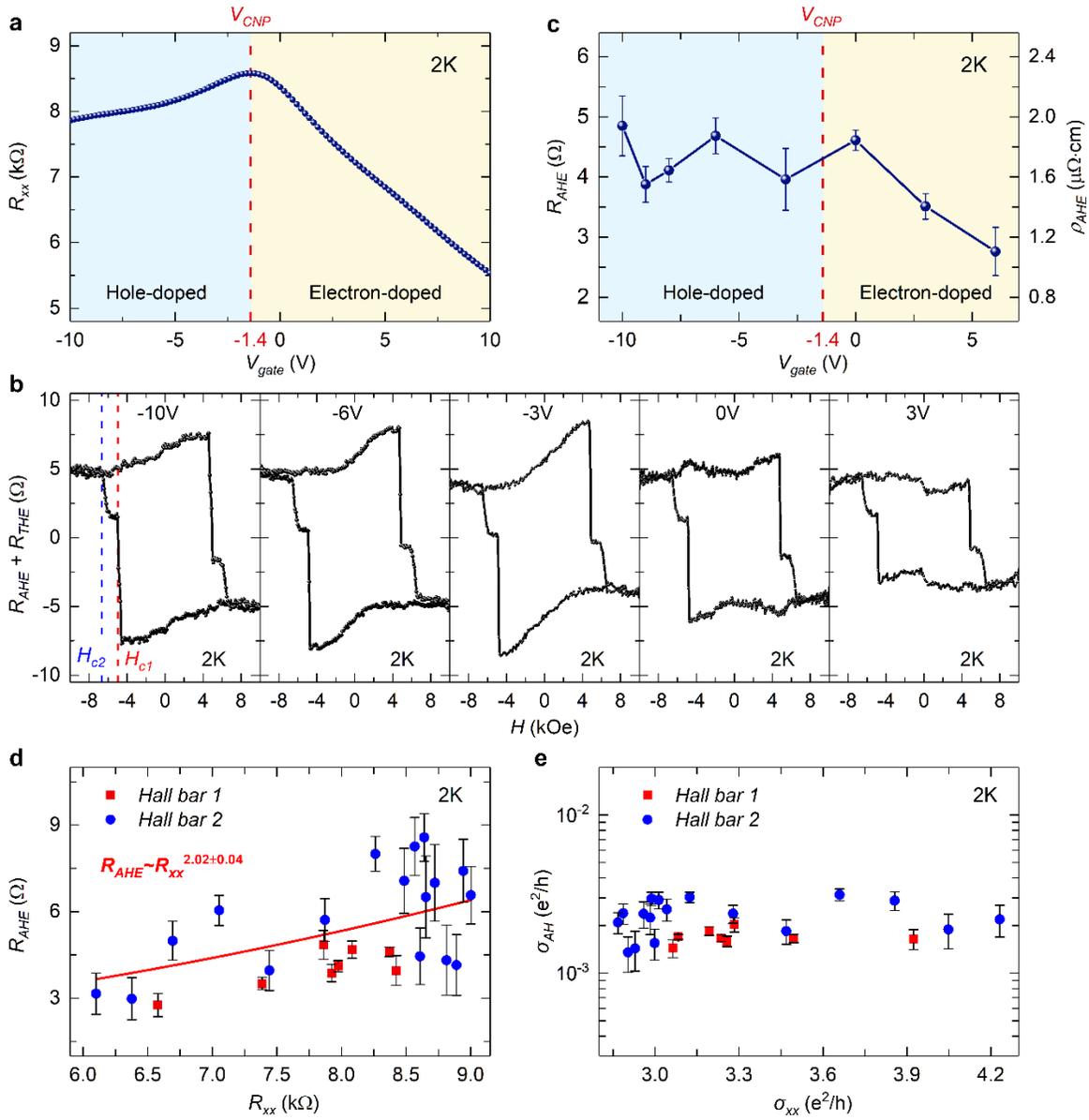

**Figure 4.** The gate dependence from -10 V to +10 V at 2 K of BST/EuIG, *sample E*. (a) $R_{xx}$. (b) AHE loops. $H_{c1}$ and $H_{c2}$ are attributed to the coercive fields of the hybridized bottom and top surfaces of BST, respectively. (c) $R_{AHE}$ (left) and $\rho_{AHE}$ (right). (d) Correlation between $R_{AHE}$ and $R_{xx}$ showing $R_{AHE} \sim R_{xx}^{2.02}$ in two Hall bars. (e) Correlation between $\sigma_{AH}$ and $\sigma_{xx}$ in the same Hall bars.

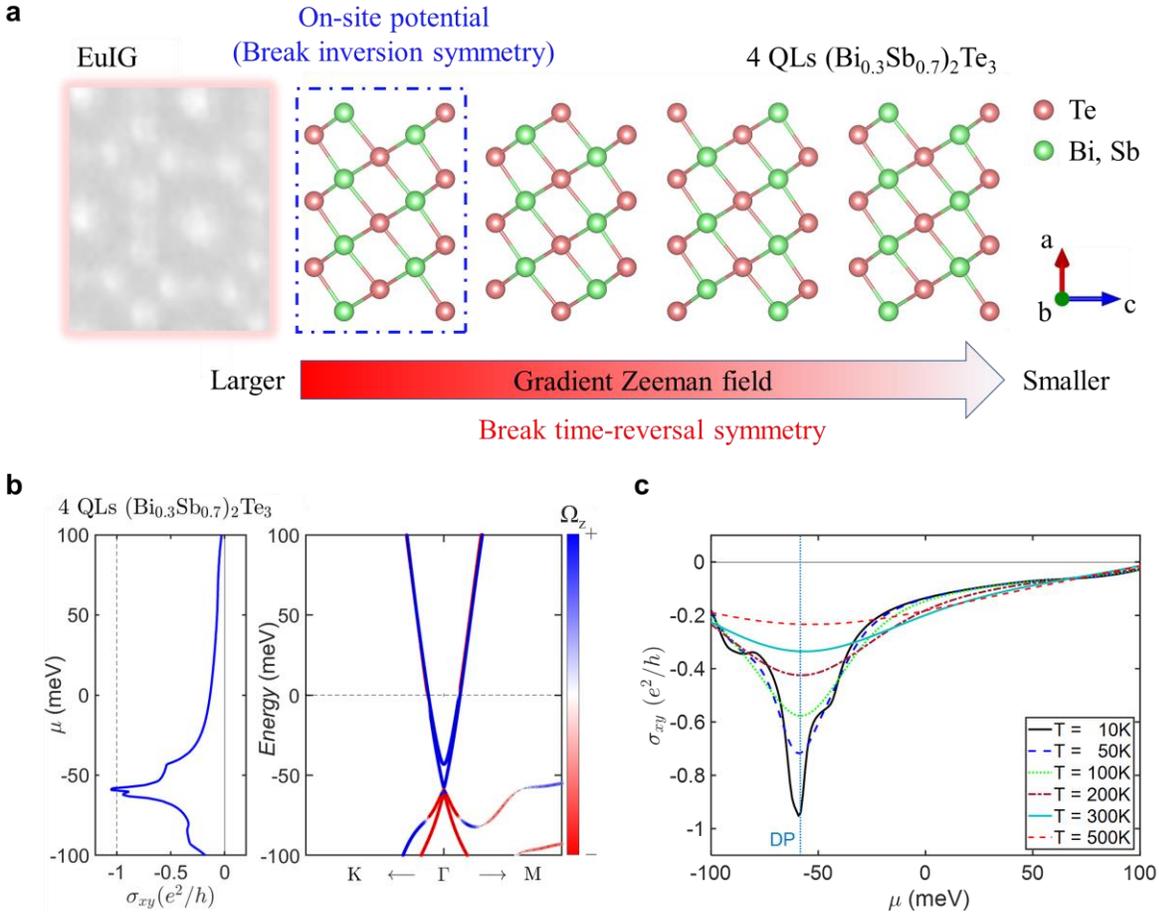

**Figure 5.** The DFT calculations of 4 QLs BST on EuIG. (a) Model of the DFT calculations in 4 QLs BST/EuIG with a gradient Zeeman field and an on-site potential. The Zeeman field values are -40 meV, -30 meV, -20 meV, and -10 meV for the 1st QL to the 4th QL of BST, respectively. The on-site potential value is 5 meV for the 1st QL of BST. (b) The calculated local BC of each band (right) and the chemical potential dependence of $\sigma_{xy}$ (left) including a gradient Zeeman field and an on-site potential at 0 K. (c) Calculated $\sigma_{xy}$ of BST *vs* chemical potential with a gradient Zeeman field and an on-site potential as a function of temperature from 10 K to 500 K.

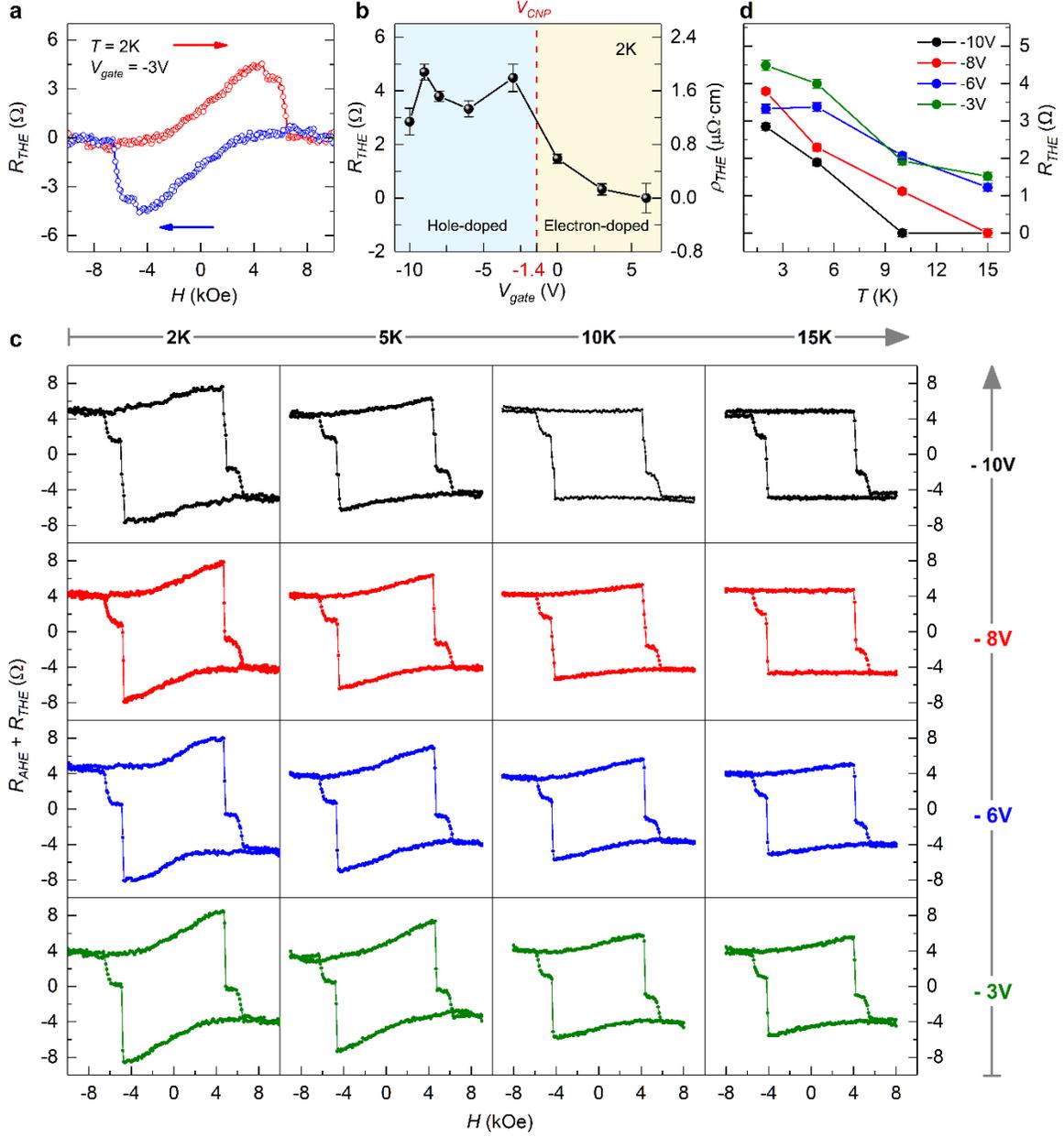

**Figure 6.** Observation of THE features in BST/EuIG. (a) The extracted $R_{THE}$ of *sample E* at -3 V and 2 K. (b) Gate dependence of $R_{THE}$ (left) and $\rho_{THE}$ (right) at 2 K. (c) Temperature dependence of AHE loops with THE. (d) $R_{THE}$ values for $V_{gate}$ at -10 V, -8 V, -6 V, and -3 V.

ASSOCIATED CONTENT

**Supporting Information**.

AFM images of EuIG films annealed at various $T_a$, XPS spectra of EuIG films annealed at various $T_a$, AFM images of BST grown on EuIG annealed at various $T_a$, suppressed weak anti-localization in BST/EuIG samples, discussion on MPE-induced and SHE-induced AHE signals, top-gate device fabrication and measurement process, gate-dependent Hall effect data, bulk topological properties of BST, the dependence of the bandgap *vs* the Zeeman field in 4 QLs and 7 QLs BST, method for extracting $R_{AHE}$ and $R_{THE}$ values, and discussion on the absence of QAHE at 2 K

AUTHOR INFORMATION

**Corresponding Authors**

* **Jueinai Kwo** – *Department of Physics, National Tsing Hua University, Hsinchu 30013, Taiwan;* orcid.org/0000-0002-5088-6677; E-mail: raynien@phys.nthu.edu.tw

* **Minghwei Hong** – *Graduate Institute of Applied Physics and Department of Physics, National Taiwan University, Taipei 10617, Taiwan;* orcid.org/0000-0003-4657-0933; E-mail: mhong@phys.ntu.edu.tw

* **Tay-Rong Chang** – *Department of Physics, National Cheng Kung University, Tainan 701, Taiwan; Center for Quantum Frontiers of Research and Technology (QFort), Tainan 701, Taiwan; Physics Division, National Center for Theoretical Sciences, National Taiwan University, Taipei 10617, Taiwan;* orcid.org/0000-0003-1222-2527; E-mail: u32trc00@phys.ncku.edu.tw

**Authors**

**Wei-Jhih Zou** – *Department of Physics, National Tsing Hua University, Hsinchu 30013, Taiwan;* orcid.org/0000-0002-8335-3711

**Meng-Xin Guo** – *Department of Physics, National Tsing Hua University, Hsinchu 30013, Taiwan;* orcid.org/0000-0001-9986-6311

**Jyun-Fong Wong** – *Department of Physics, National Tsing Hua University, Hsinchu 30013, Taiwan;* orcid.org/0000-0001-5756-5801

**Zih-Ping Huang** – *Graduate Institute of Applied Physics and Department of Physics, National Taiwan University, Taipei 10617, Taiwan;* orcid.org/0000-0001-8585-8812

**Jui-Min Chia** – *Department of Physics, National Tsing Hua University, Hsinchu 30013, Taiwan;* orcid.org/0000-0003-1759-4725

**Wei-Nien Chen** – *Department of Physics, National Tsing Hua University, Hsinchu 30013, Taiwan;* orcid.org/0000-0002-2188-0102

**Sheng-Xin Wang** – *Department of Physics, National Tsing Hua University, Hsinchu 30013, Taiwan;* orcid.org/0000-0001-9725-2930

**Keng-Yung Lin** – *Graduate Institute of Applied Physics and Department of Physics, National Taiwan University, Taipei 10617, Taiwan;* orcid.org/0000-0003-4679-7046

**Lawrence Boyu Young** – *Graduate Institute of Applied Physics and Department of Physics, National Taiwan University, Taipei 10617, Taiwan;* orcid.org/0000-0003-2569-6094

**Yen-Hsun Glen Lin** – *Graduate Institute of Applied Physics and Department of Physics, National Taiwan University, Taipei 10617, Taiwan;* orcid.org/0000-0002-0757-4109

**Mohammad Yahyavi** – *Department of Physics, National Cheng Kung University, Tainan 701, Taiwan;* orcid.org/0000-0003-0062-203X

**Chien-Ting Wu** – *Materials Analysis Division, Taiwan Semiconductor Research Institute, National Applied Research Laboratories, Hsinchu 300091, Taiwan*

**Horng-Tay Jeng** – *Department of Physics, National Tsing Hua University, Hsinchu 30013, Taiwan; Physics Division, National Center for Theoretical Sciences, National Taiwan University, Taipei 10617, Taiwan; Institute of Physics, Academia Sinica, Taipei 11529, Taiwan;* orcid.org/0000-0002-2881-3826

**Shang-Fan Lee** – *Institute of Physics, Academia Sinica, Taipei 11529, Taiwan;* orcid.org/0000-0001-5899-7200

**Author Contributions**

W.-J. Z., J.-F. W., and J.-M. C. collected the transport data and analyzed the data. M.-X. G., Z.-P. H., S.-X. W., L.-B. Y., and Y.-H. G. L. fabricated the samples. W.-N. C. performed the XRD measurements. K.-Y. L. performed the XPS measurements. M. Y. did the theoretical calculations. C.-T. W. performed the STEM measurements. T.-R. C. and S.-F. L. provided scientific supports. J. K. and M. H. supervised the project. W.-J. Z. and J.-F. W. wrote the manuscript with the comments of all the authors.

**Author Contributions**

[#]W.-J. Z., M.-X. G., J.-F. W., Z.-P. H., and J.-M. C. contributed equally to this work.

**Notes**

The authors declare no competing financial interest.

ACKNOWLEDGMENT

The authors would like to thank helpful discussions with C.-Y. Mou, H.-H. Lin, C.-F. Pai, C.-H. Hsu, Y.-C. Liu, C.-C. Chen, K.-H. M. Chen, J.-K. Cheng, and S.-W. Huang. Technical supports from NGPL/IOP/Academia Sinica, NSRRC, TSRI, and NTU Consortium of Electron Microscopy Key Technology, Taiwan, are acknowledged. This work was financially supported by the Ministry of Science and Technology (MOST), Taiwan, with project numbers 105-2112-M-007-014-MY3, 109-2112-M-002-028, 109-2622-8-002-003, Center of Quantum Technology, NTHU with Project number XXX109B0022I4, and the Thematic Project AS-TP-107-M04, Academia Sinica, Taiwan. T.-R. Chang was supported by the Young Scholar Fellowship Program from the MOST in Taiwan, under a MOST grant for the Columbus Program, No. MOST110-2636-M-006-016, NCKU, Taiwan, and National Center for Theoretical Sciences, Taiwan. Work at NCKU was supported by the MOST, Taiwan, under Grant No. MOST107-2627-E-006-001 and Higher Education Sprout Project, Ministry of Education to the Headquarters of University Advancement at NCKU.

REFERENCES

(1) Qi X.-L.; Hughes T. L.; Zhang S.-C. Topological Field Theory of Time-Reversal Invariant Insulators. *Phys. Rev. B* **2008**, *78*, 195424.

(2) Hasan M. Z.; Kane C. L. Colloquium: Topological Insulators. *Rev. Mod. Phys.* **2010**, *82*, 3045.

(3) Yu R.; Zhang W.; Zhang H.-J.; Zhang S.-C.; Dai X.; Fang Z. Quantized Anomalous Hall Effect in Magnetic Topological Insulators. *Science* **2010**, *329*, 61–64.

(4) Chang, C.-Z.; Zhang, J.; Feng, X.; Shen, J.; Zhang, Z.; Guo, M.; Li, K.; Ou, Y.; Wei, P.; Wang, L.-L.; Ji, Z.-Q.; Feng, Y.; Ji, S.; Chen, X.; Jia, J.; Dai, X.; Fang, Z.; Zhang, S.-C.; He, K.; Wang, Y.; et al. Experimental Observation of the Quantum Anomalous Hall Effect in a Magnetic Topological Insulator. *Science* **2013**, *340*, 167–170.

(5) Tokura, Y.; Yasuda, K.; Tsukazaki, A. Magnetic Topological Insulators. *Nat. Rev. Phys.* **2019**, *1*, 126–143.

(6) Kou, X.; Guo, S.-T.; Fan, Y.; Pan, L.; Lang, M.; Jiang, Y.; Shao, Q.; Nie, T.; Murata, K.; Tang, J.; Wang, Y.; He, L.; Lee, T.-K.; Lee, W.-L.; Wang, K. L. Scale-Invariant Quantum

Anomalous Hall Effect in Magnetic Topological Insulators beyond the Two-Dimensional Limit. *Phys. Rev. Lett.* **2014**, *113*, 137201.

(7) Checkelsky, J. G.; Yoshimi, R.; Tsukazaki, A.; Takahashi, K. S.; Kozuka, Y.; Falson, J.; Kawasaki, M.; Tokura, Y. Trajectory of the Anomalous Hall Effect towards the Quantized State in a Ferromagnetic Topological Insulator. *Nat. Phys.* **2014**, *10*, 731–736.

(8) Chang, C.-Z.; Zhao, W.; Kim, D. Y.; Zhang, H.; Assaf, B. A.; Heiman, D.; Zhang, S.-C.; Liu, C.; Chan, M. H. W.; Moodera, J. S. High-Precision Realization of Robust Quantum Anomalous Hall State in a Hard Ferromagnetic Topological Insulator. *Nat. Mater.* **2015**, *14*, 473–477.

(9) Kandala, A.; Richardella, A.; Kempinger, S.; Liu, C.-X.; Samarth, N. Giant Anisotropic Magnetoresistance in a Quantum Anomalous Hall Insulator. *Nat. Commun.* **2015**, *6*, 7434.

(10) Mogi, M.; Yoshimi, R.; Tsukazaki, A.; Yasuda, K.; Kozuka, Y.; Takahashi, K. S.; Kawasaki, M.; Tokura, Y. Magnetic Modulation Doping in Topological Insulators toward Higher-Temperature Quantum Anomalous Hall Effect. *Appl. Phys. Lett.* **2015**, *107*, 182401.

(11) Chang, C.-Z.; Tang, P.; Wang, Y.-L.; Feng, X.; Li, K.; Zhang, Z.; Wang, Y.; Wang, L.-L.; Chen, X.; Liu, C.; Duan, W.; He, K.; Ma, X.-C.; Xue, Q.-K. Chemical-Potential-Dependent Gap Opening at the Dirac Surface States of $Bi_2Se_3$ Induced by Aggregated Substitutional Cr Atoms. *Phys. Rev. Lett.* **2014**, *112*, 056801.

(12)   Wei, P.; Katmis, F.; Assaf, B. A.; Steinberg, H.; Jarillo-Herrero, P.; Heiman, D.; Moodera, J. S. Exchange-Coupling-Induced Symmetry Breaking in Topological Insulators. *Phys. Rev. Lett.* **2013**, *110*, 186807.

(13)   Katmis, F.; Lauter, V.; Nogueira, F. S.; Assaf, B. A.; Jamer, M. E.; Wei, P.; Satpati, B.; Freeland, J. W.; Eremin, I.; Heiman, D.; Jarillo-Herrero, P.; Moodera, J. S. A High-Temperature Ferromagnetic Topological Insulating Phase by Proximity Coupling. *Nature* **2016**, *533*, 513–516.

(14)   Lang, M.; Montazeri, M.; Onbasli, M. C.; Kou, X.; Fan, Y.; Upadhyaya, P.; Yao, K.; Liu, F.; Jiang, Y.; Jiang, W.; Wong, K. L.; Yu, G.; Tang, J.; Nie, T.; He, L.; Schwartz, R. N.; Wang, Y.; Ross, C. A.; Wang, K. L. Proximity Induced High-Temperature Magnetic Order in Topological Insulator - Ferrimagnetic Insulator Heterostructure. *Nano Lett.* **2014**, *14*, 3459–3465.

(15)    Bhattacharyya, S.; Akhgar, G.; Gebert, M.; Karel, J.; Edmonds, M. T.; Fuhrer, M. S. Recent Progress in Proximity Coupling of Magnetism to Topological Insulators. *Adv. Mater.* **2021**, *33*, e2007795.

(16)    Tang, C.; Chang, C.-Z.; Zhao, G.; Liu, Y.; Jiang, Z.; Liu, C.-X.; McCartney, M. R.; Smith, D. J.; Chen, T.; Moodera, J. S.; Shi, J. Above 400-K Robust Perpendicular Ferromagnetic Phase in a Topological Insulator. *Sci. Adv.* **2017**, *3*, e1700307.

(17)    Chen, C. C.; Chen, K. H. M.; Fanchiang, Y. T.; Tseng, C. C.; Yang, S. R.; Wu, C. N.; Guo, M. X.; Cheng, C. K.; Huang, S. W.; Lin, K. Y.; Wu, C. T.; Hong, M.; Kwo, J. Topological Insulator $Bi_2Se_3$ Films on Rare Earth Iron Garnets and Their High-Quality Interfaces. *Appl. Phys. Lett.* **2019**, *114*, 031601.

(18)    Liu, Y. C.; Chen, C. C.; Fanchiang, Y. T.; Yang, S. R.; Zou, W. J.; Guo, M. X.; Cheng, C. K.; Chen, K. H. M.; Lee, S. F.; Hong, M.; Kwo, J. Features of Magnetic Proximity Effect in the Magneto-Transport of $(Bi,Sb)_2Te_3$ on TmIG. unpublished.

(19)    Tang, C.; Sellappan, P.; Liu, Y.; Xu, Y.; Garay, J. E.; Shi, J. Anomalous Hall Hysteresis in $Tm_3Fe_5O_{12}$/Pt with Strain-Induced Perpendicular Magnetic Anisotropy. *Phys. Rev. B* **2016**, *94*, 140403.

(20)    Avci, C. O.; Quindeau, A.; Pai, C.-F.; Mann, M.; Caretta, L.; Tang, A. S.; Onbasli, M. C.; Ross, C. A.; Beach, G. S. D. Current-Induced Switching in a Magnetic Insulator. *Nat. Mater.* **2017**, *16*, 309–314.

(21)    Wu, C. N.; Tseng, C. C.; Lin, K. Y.; Cheng, C. K.; Yeh, S. L.; Fanchiang, Y. T.; Hong, M.; Kwo, J. High-Quality Single-Crystal Thulium Iron Garnet Films with Perpendicular Magnetic Anisotropy by Off-Axis Sputtering. *AIP Advances* **2018**, *8*, 055904.

(22)    Wu, C. N.; Tseng, C. C.; Fanchiang, Y. T.; Cheng, C. K.; Lin, K. Y.; Yeh, S. L.; Yang, S. R.; Wu, C. T.; Liu, T.; Wu, M.; Hong, M.; Kwo, J. High-Quality Thulium Iron Garnet Films with Tunable Perpendicular Magnetic Anisotropy by Off-Axis Sputtering – Correlation between Magnetic Properties and Film Strain. *Sci. Rep.* **2018**, *8*, 11087.

(23)    Rosenberg, E. R.; Beran, L.; Avci, C. O.; Zeledon, C.; Song, B.; Gonzalez-Fuentes, C.; Mendil, J.; Gambardella, P.; Veis, M.; Garcia, C.; Beach, G. S. D.; Ross, C. A. Magnetism and Spin Transport in Rare-Earth-Rich Epitaxial Terbium and Europium Iron Garnet Films. *Phys. Rev. Mater.* **2018**, *2*, 094405.

(24)     Ortiz, V. H.; Aldosary, M.; Li, J.; Xu, Y.; Lohmann, M. I.; Sellappan, P.; Kodera, Y.; Garay, J. E.; Shi, J. Systematic Control of Strain-Induced Perpendicular Magnetic Anisotropy in Epitaxial Europium and Terbium Iron Garnet Thin Films. *APL Mater.* **2018**, *6*, 121113.

(25) Guo, M. X.; Liu, Y. C.; Wu, C. N.; Cheng, C. K.; Chen, W. N.; Chen, T. Y.; Kuo, C. Y.; Chang, C. F.; Tjeng, L. H.; Zhou, S. Q.; Pai, C. F.; Hong, M.; Kwo, J. Single-Crystal Europium Iron Garnet Films with Perpendicular Magnetic Anisotropy by Sputtering. unpublished.

(26) Nagaosa, N.; Tokura, Y. Topological Properties and Dynamics of Magnetic Skyrmions. *Nat. Nanotechnol.* **2013**, *8*, 899–911.

(27) Li, P.; Ding, J.; Zhang, S. S.-L.; Kally, J.; Pillsbury, T.; Heinonen, O. G.; Rimal, G.; Bi, C.; DeMann, A.; Field, S. B.; Wang, W.; Tang, J.; Jiang, J. S.; Hoffmann, A.; Samarth, N.; Wu, M. Topological Hall Effect in a Topological Insulator Interfaced with a Magnetic Insulator. *Nano Lett.* **2021**, *21*, 84–90.

(28) Fert, A.; Cros, V.; Sampaio, J. Skyrmions on the Track. *Nat. Nanotechnol.* **2013**, *8*, 152–156.

(29) Iida, S. Magnetostriction Constants of Rare Earth Iron Garnets. *J. Phys. Soc. Japan.* **1967**, *22*, 1201–1209.

(30)     Zanjani, S. M.; Onbaşlı, M. C. Predicting New Iron Garnet Thin Films with Perpendicular Magnetic Anisotropy. *J. Magn. Magn.* **2020**, *499*, 166108.

(31) Watanabe, R.; Yoshimi, R.; Kawamura, M.; Mogi, M.; Tsukazaki, A.; Yu, X. Z.; Nakajima, K.; Takahashi, K. S.; Kawasaki, M.; Tokura, Y. Quantum Anomalous Hall Effect Driven by Magnetic Proximity Coupling in All-Telluride Based Heterostructure. *Appl. Phys. Lett.* **2019**, *115*, 102403.

(32) Chang, C.-Z.; Zhang, J.; Liu, M.; Zhang, Z.; Feng, X.; Li, K.; Wang, L.-L.; Chen, X.; Dai, X.; Fang, Z.; Qi, X.-L.; Zhang, S.-C.; Wang, Y.; He, K.; Ma, X.-C.; Xue, Q.-K. Thin Films of Magnetically Doped Topological Insulator with Carrier-Independent Long-Range Ferromagnetic Order. *Adv. Mater.* **2013**, *25*, 1065.

(33) Liu, C.; Zang, Y.; Ruan, W.; Gong, Y.; He, K.; Ma, X.; Xue, Q.-K.; Wang, Y. Dimensional Crossover-Induced Topological Hall Effect in a Magnetic Topological Insulator. *Phys. Rev. Lett.* **2017**, *119*, 176809.

(34) Jiang, J.; Xiao, D.; Wang, F.; Shin, J.-H.; Andreoli, D.; Zhang, J.; Xiao, R.; Zhao, Y.-F.; Kayyalha, M.; Zhang, L.; Wang, K.; Zang, J.; Liu, C.; Samarth, N.; Chan, M. H. W.; Chang, C.-

Z. Concurrence of Quantum Anomalous Hall and Topological Hall Effects in Magnetic Topological Insulator Sandwich Heterostructures. *Nat. Mater.* **2020**, *19*, 732.

(35) Hao, Q.; Xiao, G. Giant Spin Hall Effect and Magnetotransport in a Ta/CoFeB/MgO Layered Structure: A Temperature Dependence Study. *Phys. Rev. B* **2015**, *91*, 224413.

(36) Shao, Q.; Grutter, A.; Liu, Y.; Yu, G.; Yang, C.-Y.; Gilbert, D. A.; Arenholz, E.; Shafer, P.; Che, X.; Tang, C.; Aldosary, M.; Navabi, A.; He, Q. L.; Kirby, B. J.; Shi, J.; Wang, K. L. Exploring Interfacial Exchange Coupling and Sublattice Effect in Heavy Metal/Ferrimagnetic Insulator Heterostructures Using Hall Measurements, X-Ray Magnetic Circular Dichroism, and Neutron Reflectometry. *Phys. Rev. B* **2019**, *99*, 104401.

(37) Nagaosa, N.; Sinova, J.; Onoda, S.; MacDonald, A. H.; Ong, N. P. Anomalous Hall Effect. *Rev. Mod. Phys.* **2010**, *82*, 1539.

(38) Jiang, Z.; Chang, C.-Z.; Tang, C.; Wei, P.; Moodera, J. S.; Shi, J. Independent Tuning of Electronic Properties and Induced Ferromagnetism in Topological Insulators with Heterostructure Approach. *Nano Lett.* **2015**, *15*, 5835–5840.

(39) Guo, G. Y.; Yao, Y.; Niu, Q.; *Ab Initio* Calculation of the Intrinsic Spin Hall Effect in Semiconductors. *Phys. Rev. Lett.* **2005**, *94*, 226601.

(40) Wang, Q.-Z.; Liu, X.; Zhang, H.-J.; Samarth, N.; Zhang, S.-C.; Liu, C.-X. Quantum Anomalous Hall Effect in Magnetically Doped InAs/GaSb Quantum Wells. *Phys. Rev. Lett.* **2014**, *113*, 147201.

(41) Gao, A.; Liu, Y.-F.; Hu, C.; Qiu, J.-X.; Tzschaschel, C.; Ghosh, B.; Ho, S.-C.; Bérubé, D.; Chen, R.; Sun, H.; Zhang, Z.; Zhang, X.-Y.; Wang, Y.-X.; Wang, N.; Huang, Z.; Felser, C.; Agarwal, A.; Ding, T.; Tien, H.-J.; Akey, A.; et al. Layer Hall Effect in a 2D Topological Axion Antiferromagnet. *Nature* **2021**, *595*, 521–525.

(42) Kresse, G.; Furthmüller, J. Efficient Iterative Schemes for *Ab Initio* Total-Energy Calculations Using a Plane-Wave Basis Set. *Phys. Rev. B* **1996**, *54*, 11169.

(43) Kresse, G.; Joubert, D. From Ultrasoft Pseudopotentials to the Projector Augmented-Wave Method. *Phys. Rev. B* **1999**, *59*, 1758.

(44) Blöchl, P. E. Projector Augmented-Wave Method. *Phys. Rev. B* **1994**, *50*, 17953.

(45) Marzari, N.; Vanderbilt, D. Maximally Localized Generalized Wannier Functions for Composite Energy Bands. *Phys. Rev. B* **1997**, *56*, 12847.

(46) Souza, I.; Marzari, N.; Vanderbilt, D.; Maximally Localized Wannier Functions for Entangled Energy Bands. *Phys. Rev. B* **2001**, *65*, 035109.

(47) Mostofi, A. A.; Yates, J. R.; Lee, Y.-S.; Souza, I.; Vanderbilt, D.; Marzari, N. Wannier90: A Tool for Obtaining Maximally-Localised Wannier Functions. *Comput Phys Commun.* **2008**, *178*, 685–699.

(48) Franchini, C.; Kováčik, R.; Marsman, M.; Murthy, S. S.; He, J.; Ederer, C.; Kresse, G. Maximally Localized Wannier Functions in LaMnO$_3$ within PBE + U, Hybrid Functionals and Partially Self-Consistent GW: an Efficient Route to Construct *Ab Initio* Tight-Binding Parameters for e$_g$ Perovskites. *J. Phys.: Condens. Matter* **2012**, *24*, 235602.

# Enormous Berry-Curvature-Driven Anomalous Hall Effect in Topological Insulator (Bi,Sb)$_2$Te$_3$ on Ferrimagnetic Europium Iron Garnet beyond 400 K

Wei-Jhih Zou[1,#], Meng-Xin Guo[1,#], Jyun-Fong Wong[1,#], Zih-Ping Huang[2,#], Jui-Min Chia[1,#], Wei-Nien Chen[1], Sheng-Xin Wang[1], Keng-Yung Lin[2], Lawrence Boyu Young[2], Yen-Hsun Glen Lin[2], Mohammad Yahyavi[3], Chien-Ting Wu[4], Horng-Tay Jeng[1,5,6], Shang-Fan Lee[5], Tay-Rong Chang[3,6,7*], Minghwei Hong[2*], Jueinai Kwo[1*]

[1]Department of Physics, National Tsing Hua University, Hsinchu 30013, Taiwan

[2]Graduate Institute of Applied Physics and Department of Physics, National Taiwan University, Taipei 10617, Taiwan

[3]Department of Physics, National Cheng Kung University, Tainan 701, Taiwan

[4]Materials Analysis Division, Taiwan Semiconductor Research Institute, National Applied Research Laboratories, Hsinchu 300091, Taiwan

[5]Institute of Physics, Academia Sinica, Taipei 11529, Taiwan

[6]Physics Division, National Center for Theoretical Sciences, National Taiwan University, Taipei 10617, Taiwan

[7]Center for Quantum Frontiers of Research and Technology (QFort), Tainan 701, Taiwan

[#]W.-J. Z., M.-X. G., J.-F. W., Z.-P. H., and J.-M. C. contributed equally to this work.

[*]Address correspondence to J. Kwo, raynien@phys.nthu.edu.tw; M. Hong, mhong@phys.ntu.edu.tw; T.-R. Chang, u32trc00@phys.ncku.edu.tw

**Contents**

**Section S1. AFM images of EuIG films annealed at various $T_a$**

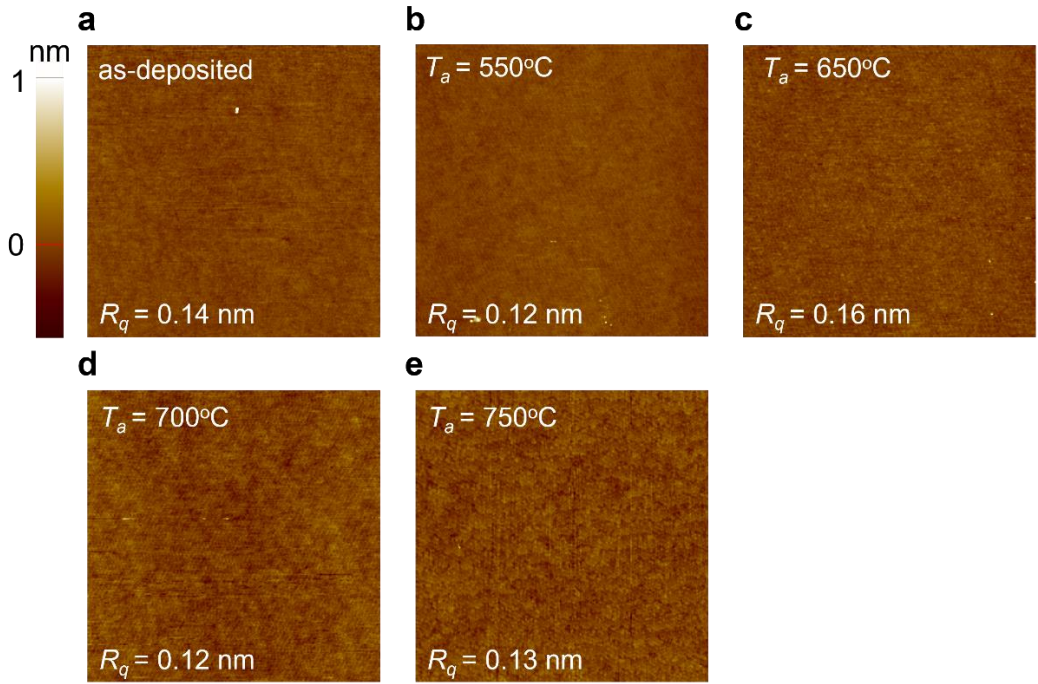

**Figure S1**. Surface morphologies (in a 5×5 μm² area) of EuIG with $Eu/Fe = 0.529$ by AFM. (a) As-deposited. (b) to (e) Annealed films at various temperatures.

**Section S2. XPS spectra of EuIG films annealed at various $T_a$**

EuIG films were annealed in an ultra-high vacuum before the MBE growth of BST to remove the surface carbon contamination caused by the *ex situ* transfer of samples from the sputtering chamber to the MBE chamber. XPS was utilized to examine the effect of $T_a$ on the surface chemistry of EuIG, using an Al $K_\alpha$ excitation source (1486.6 eV) and an electron flood gun for charge neutralization. Figure S2a shows the C 1*s* core-level spectrum of as-deposited EuIG exhibiting pronounced C signals due to contamination during air exposure; in contrast, much reduced C peaks in the annealed samples indicate that the contamination was mostly removed.

Figure S2b shows the Eu 3*d* and Fe 2*p* core-level spectra of annealed EuIG films, showing the increasing Fe areal intensity relative to that of Eu as the annealing temperature was elevated from 550 °C to 750 °C (inset of Figure S2b).

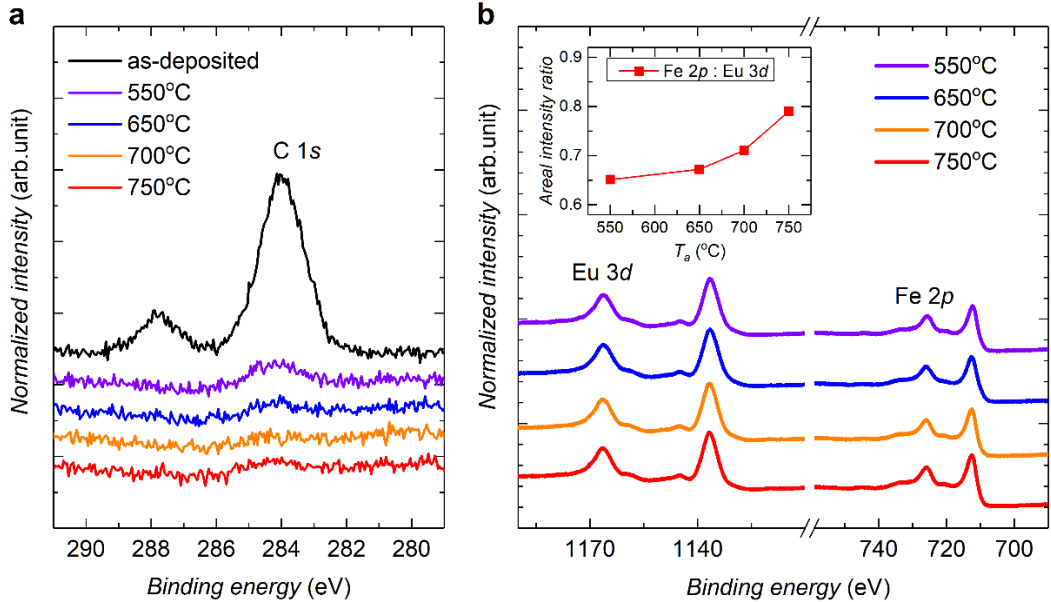

**Figure S2.** XPS analysis of bare EuIG. (a) XPS C 1*s*. (b) Eu 3*d* and Fe 2*p* core-level spectra of as-deposited EuIG film and films annealed to various temperatures. For a EuIG film annealed to higher temperatures, the C contamination was reduced drastically, and the *Eu/Fe* was decreased.

**Section S3. AFM images of BST grown on EuIG annealed at various $T_a$**

To investigate the impact of the EuIG surface condition on the growth of BST, the $T_a$ of EuIG was varied from 450 °C to 750 °C to optimize the starting surface and compared the quality of BST thin films grown on. Surface morphologies of 7 nm thick BST films by AFM are displayed in Figure S3. The smoothest surface with the smallest surface roughness $R_q$ of 0.65 nm and terraced triangular domains were observed for $T_a$ of 650 °C, illustrating the layer-by-layer growth of BST. In addition, the visually increasing of the domain size indicates that the vdW epitaxy was promoted when $T_a$ was increased. However, some cracks were observed when EuIG was annealed exceeding 750 °C, and thus the optimal range of $T_a$ was between 650 °C and 700 °C.

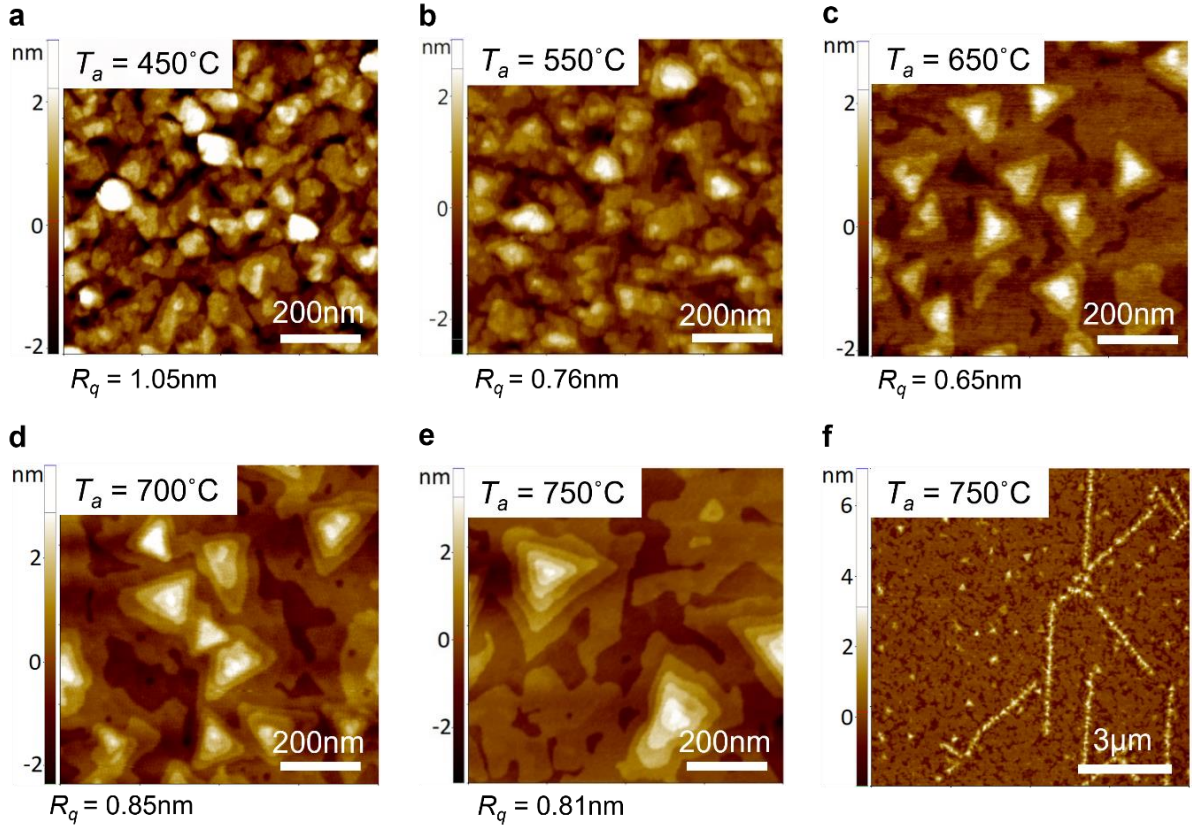

**Figure S3.** Surface morphologies (in a 1×1 μm² area) by AFM for 7 nm BST grown on EuIG with *Eu/Fe* = 0.529 and $T_a$ of (a) 450 °C. (b) 550 °C. (c) 650 °C. (d) 700 °C. (e) 750 °C. (f) The cracked BST film surface on EuIG with *Eu/Fe* = 0.577 and $T_a$ of 750 °C.

**Section S4. Suppressed weak anti-localization in BST/EuIG samples**

Figure S4a shows the magneto-conductance of *samples A, B, C,* and *D* at 10 K, showing weak anti-localization (WAL) behavior that can be described by the HLN equation, $\Delta\sigma_{xx} = \sigma_{xx}(B) - \sigma_{xx}(0) = \frac{\alpha e^2}{2\pi^2 \hbar} ln\left(\frac{\hbar}{4eL_\phi^2 B}\right) - \psi\left(\frac{1}{2} + \frac{\hbar}{4eL_\phi^2 B}\right)$, where $\psi$ is the digamma function and $L_\phi$ is the phase coherence length.[1] The pre-factor $α$ is theoretically expected to be -0.5 for an independent coherence channel. The parameters extracted from the HLN equation at low magnetic fields for four BST/EuIG samples are listed in Table S1. Besides, the $α$ values are marked in Figure S4b, and all of them are substantially less than -1.0. This finding indicates that the WAL of the bottom conducting channel is suppressed by the WL, as the bottom SS opens an exchange gap at the DP. Gradually suppressed WAL in these four samples comes with larger $R_{AHE}$ values, clearly shown in Figure S4b. In addition, the phase coherence length $L_\phi$ derived from the HLN equation decreases with the increased $R_{AHE}$, which could also be regarded as an enhanced exchange coupling strength.

The modulation of the $R_{AHE}$ *via* the external electric field of the top gate is shown in Figure 4c, and it is interesting that the WAL is suppressed with the carrier changing from electrons to holes (Figure S4c). As shown in Figure 4, while we apply a negative $V_{gate}$, the $E_F$ of the top SS gradually stays away from the DP of the top SS, and the $E_F$ of the bottom SS moves toward the center of the exchange gap, both resulting in the suppressed WAL. Furthermore, Figure S4d shows the positive correlation between $R_{AHE}$ and $α$ extracted from the gate-dependent magneto-conductance of *sample E*, and it shows a similar trend with the results demonstrated in Figure S4b.

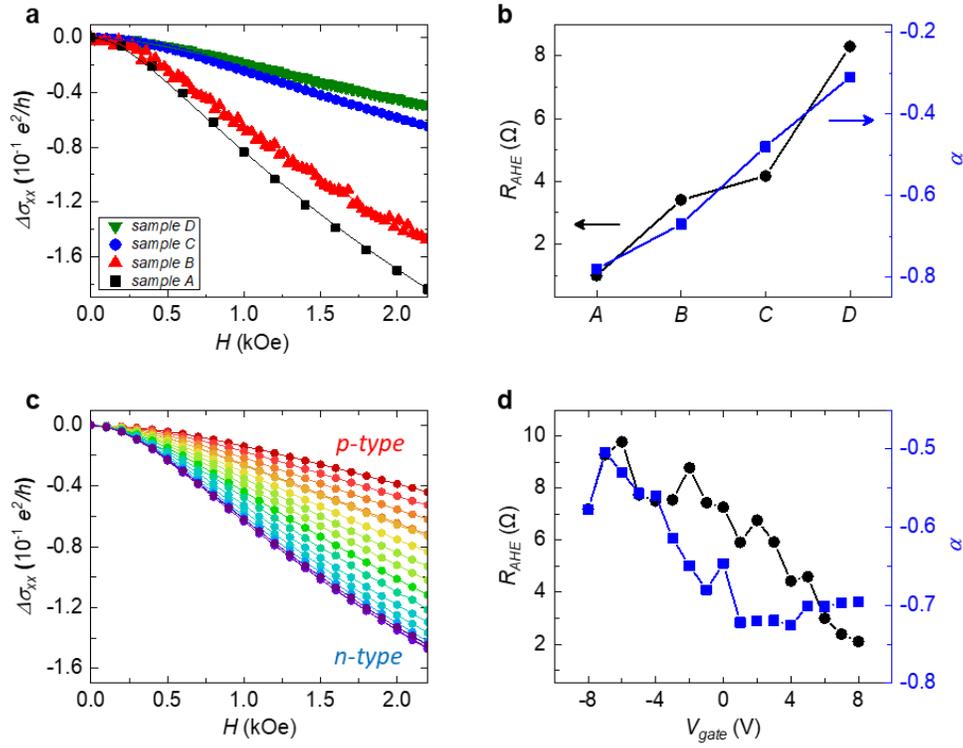

**Figure S4.** Analysis of suppressed WAL in BST/EuIG. (a) Magneto-conductance of four BST/EuIG *samples A*, *B*, *C*, and *D* measured at 10 K showing WAL behaviors. The scattered points and the solid lines are the raw data and the well-fitted curves of the HLN equation, respectively. (b) $R_{AHE}$ values at 300 K and α values for the four samples extracted from the HLN equation show a strong positive correlation. (c) $V_{gate}$ dependent magneto-conductance of BST/EuIG *sample E* of the top-gate device in Figure 4 measured at 2 K. The gradually suppressed WAL behavior was observed from the *n*-type to *p*-type region. Details of the top-gate device are given in Section 6, Supporting Information. (d) The gate-dependent $R_{AHE}$ values and *α* values of *sample E* at 2 K. It also shows a positive correlation.

**Table S1.** Magneto-conductance fitting parameters extracted by the HLN equation.

|  | BST thickness [nm] | $\alpha$ | $L_\phi$ [nm] | $R_{AHE}$ [$\Omega$, at 300 K] |
|---|---|---|---|---|
| *sample A* | 6.7 | -0.78 | 84.81 | 1.00 |
| *sample B* | 5.9 | -0.67 | 81.92 | 3.41 |
| *sample C* | 6.5 | -0.48 | 63.97 | 4.17 |
| *sample D* | 4.1 | -0.31 | 68.17 | 8.29 |

**Section S5. Discussion on MPE-induced and SHE-induced AHE signals**

Figure S5a shows the MR of $R_\perp$, $R_T$, and $R_\parallel$ for BST/EuIG, where $MR = \frac{R(\mu_0 H) - R(0)}{R(0)} \times 100\%$. MR of $R_\perp$ has a stronger relationship with the external applied magnetic field than those of $R_T$ and $R_\parallel$ due to Kohler's rule, $\frac{\Delta \rho_\perp}{\rho} = \left(\frac{R_H}{\rho}\right)^2 B_\perp^2$, where $R_H$ is the Hall coefficient. As discussed in Figure 4a, $E_F$ of the BST samples in this work is close to the DP indicating a large $R_H$ value, and therefore $\nabla_H R_\perp \gg \nabla_H R_\parallel, \nabla_H R_T$. Moreover, when the current is parallel to the magnetic field, the electronic orbits are perpendicular to the current as well as the magnetic field, which contributes to the increased scattering cross-section, leading to a higher resistance. On the contrary, a lower resistance will be found when the current is perpendicular to the in-plane magnetic field. Hence, $\nabla_H R_\perp \gg \nabla_H R_\parallel > \nabla_H R_T$ in BST/EuIG was observed.

On the other hand, the relationship between MR of $R_\perp$, $R_T$, and $R_\parallel$ for Pt/EuIG are poles apart from that of BST/EuIG. Note that, it is commonly recognized that the AHE in a heavy metal/FI heterostructure results from SHE. Figure S5b shows the MR for Pt/EuIG, where $R_T$ has a stronger dependence on the external applied magnetic field than those of $R_\perp$ and $R_{||}$, yielding $\nabla_H R_T \gg \nabla_H R_{||} \sim \nabla_H R_\perp$. Based on this relationship between MR, we could rule out SHE to be the dominant source of the AHE in BST/EuIG samples.

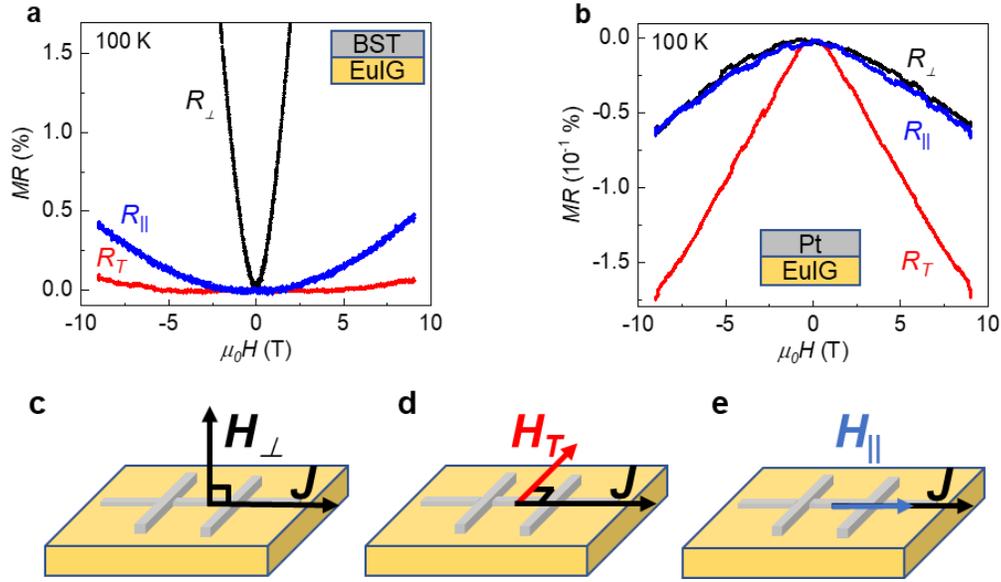

**Figure S5.** Angle-dependent MR. (a) and (b) show the magneto-resistance ratio in three magnetic fields applied directions for BST and Pt on EuIG, respectively. The directions of applied magnetic fields and their corresponding notations are illustrated in (c) to (e).

### Section S6. Top-gate device fabrication and measurement process

*Sample E* was fabricated from s*ample D* (4 nm thick BST/EuIG) using top-gate processing, and the device structure is shown in Figure S6. The gate oxide is made of a combination of $Y_2O_3$ (2 nm) and $Al_2O_3$ (15 nm), *in situ* grown by MBE and atomic layer deposition (ALD), respectively, in an integrated multi-chamber ultra-high vacuum (UHV) system, respectively. A second oxide layer made of $Al_2O_3$ (25 nm) is deposited by a separate ALD system over the entire substrate area to avoid leakages at the edges of the Hall bar.

For a better signal-to-noise ratio, the devices were measured with a standard lock-in technique at a low frequency (∼23 Hz) and a 1 μA alternating current in the PPMS.

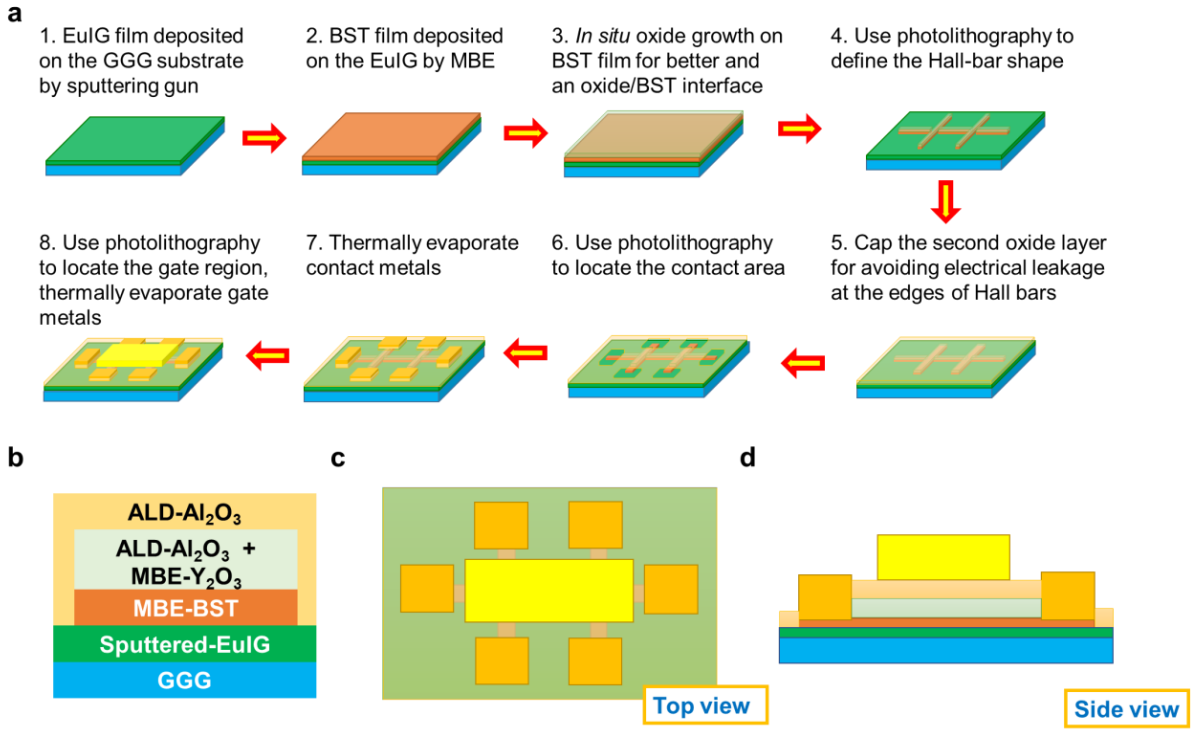

**Figure S6.** Schematic diagram of the top-gate device. (a) Fabrication process. (b) Top-gate device structure. (c) and (d) From the top view and side view, respectively.

### Section S7. Gate-dependent Hall effect data

Figure S7a shows the Hall effect of *sample E*, *Hall Bar 2* with varying $V_{gate}$. The slope changed from positive to negative with $V_{gate}$ from -8 V to 8 V, indicating the ambipolar transport behavior. Moreover, small humps manifested in AHE loops at $V_{gate}$ from -8 V to 3 V, which was a signature of the THE. This hump feature is more pronounced for a larger negative $V_{gate}$. Figure S7b shows the AHE loops of *sample E*, *Hall Bar 2* at -8 V. There are two $H_c$'s ($H_{c1}$ and $H_{c2}$) that appeared in all AHE loops. The $H_{c1}$ defined here is attributed to the "large jump" of $R_{AHE}$. The larger jump is the $H_{c1}$ of the bottom layer, and the smaller one is the $H_{c2}$ of the top layer. The $H_{c1}$ is larger than $H_{c2}$ in *Hall bar 2*, which is contrary to *Hall bar 1* ($H_{c2} > H_{c1}$). The disorder could be the cause to flip these two coercive fields. The bottom layer seems to have the stronger disorder. Moreover, this also reflects the decreasing of the step (*i.e.*, |$H_{c2}$ - $H_{c1}$|), and it may just be a feature only observed in *sample E*, *Hall bar 2*.

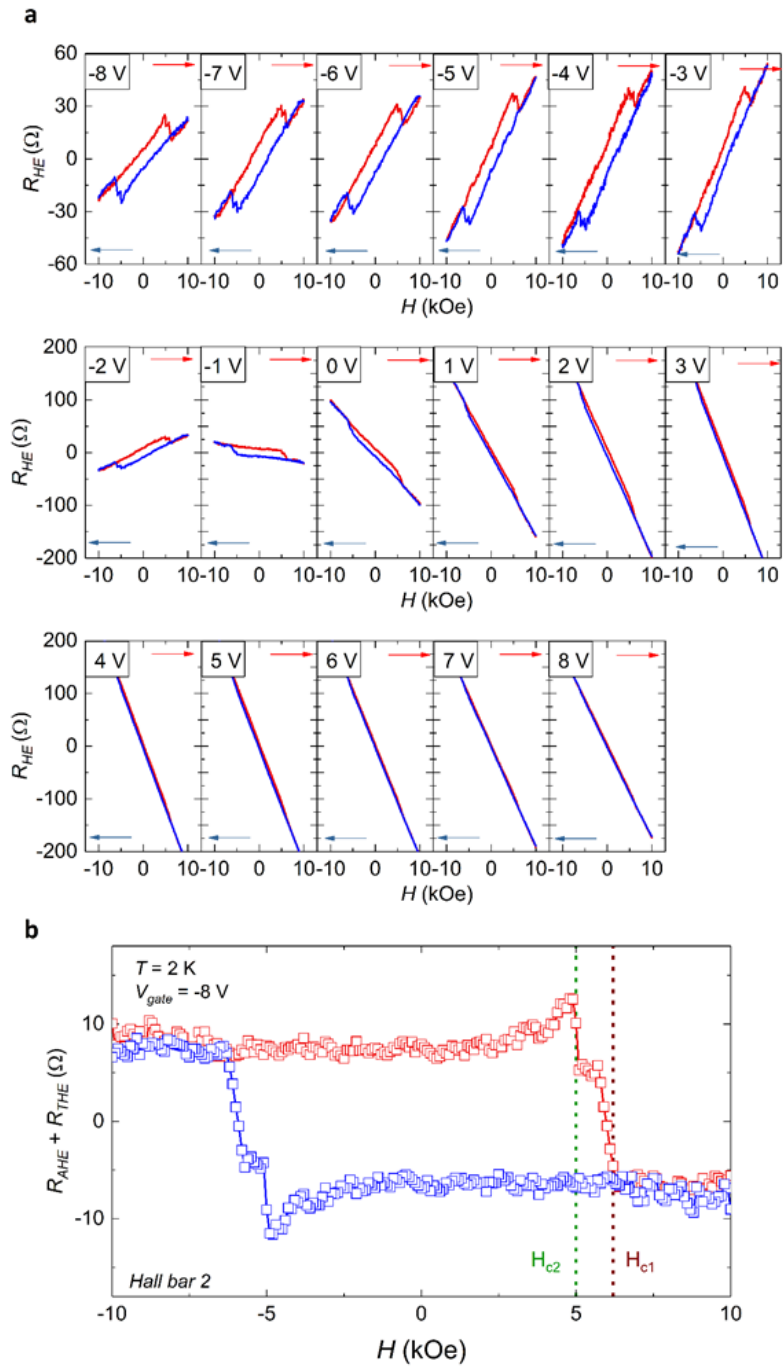

**Figure S7.** Hall effect of the top-gate device (*sample E*), *Hall bar 2* at 2 K. (a) Hall effect with $V_{gate}$ from -8 V to 8 V. (b) AHE loop at -8 V.

**Section S8. Bulk topological properties of BST**

Bi$_2$Te$_3$, Sb$_2$Te$_3$, and BST crystallizes in a rhombohedral structure with a space group $D_{3d}^5$ $R\bar{3}m$. The basis consists of five atoms in one unit cell. It can be seen that this crystal structure has space-inversion symmetry. Bi$_2$Te$_3$, Sb$_2$Te$_3$, and BST can practically be handled in units of quintuple layers. Bi$_2$Te$_3$ and Sb$_2$Te$_3$ were recently found to be three-dimensional strong topological insulators.[2] To gain additional insight into the topological properties of BST, from constructed tight-binding Hamiltonian with Bi:Sb = 3:7, we investigated the band topology. Figure S8a–c show the band structure of Bi$_2$Te$_3$, Sb$_2$Te$_3$, and BST with SOC, respectively. To firmly establish the topological properties ($Z_2$ invariant) of these materials, we followed the Wilson loop method for determining topological properties in the topological insulator (not shown).[3–5] As expected, we found these materials are strong topological insulators. Our results are consistent with the previous calculations.[2] We further calculated the surface-state energy spectra for the semi-infinite boundary by constructing the iterative Green's function from the real-space tight-binding Hamiltonian.[6] Then, using the imaginary part of the iterative Green's function or the surface Green's function ($G_s$), we calculated the local density of states at the surface $A(k,\omega) = -\frac{1}{\pi}\lim_{\eta\to 0^+} Im\, Tr\, G_s(k, \omega + i\eta)$, and according to Figure S8d–f, a comparison of the SS energy dispersion calculations for Bi$_2$Te$_3$, Sb$_2$Te$_3$, and BST demonstrated that a single topologically protected SS is located in the bulk energy gap around Γ point. The features of topological properties of BST are similar to Bi$_2$Te$_3$, Sb$_2$Te$_3$.[2]

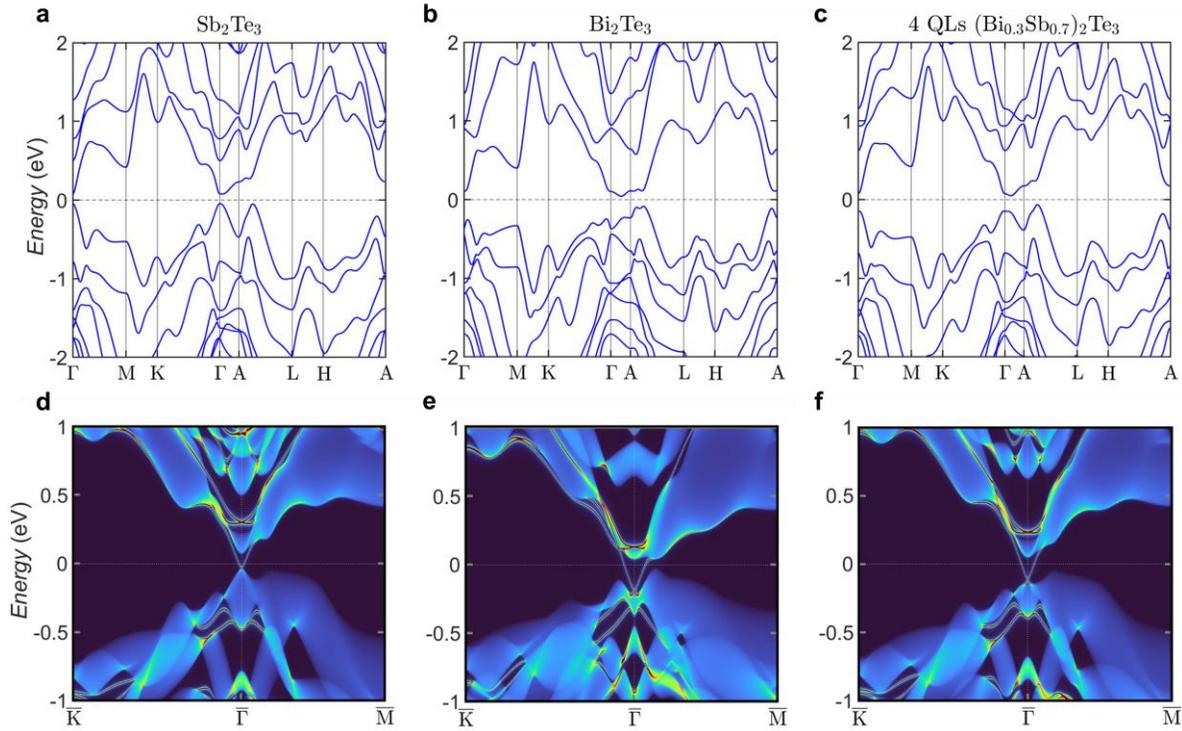

**Figure S8.** Bulk band structures of $Bi_2Te_3$, $Sb_2Te_3$, and BST. Band structure calculations with SOC along high-symmetry lines of (a) $Sb_2Te_3$, (b) $Bi_2Te_3$, and (c) BST. The dashed line indicates the $E_F$. The calculated SSs energy dispersion of (d) $Sb_2Te_3$, (e) $Bi_2Te_3$, and (f) BST on the [111] surface for the semi-infinite boundary by constructing the iterative Green's function from the real-space tight-binding Hamiltonian.[6] The SSs for $Bi_2Te_3$, $Sb_2Te_3$, and BST can be clearly seen around the Γ point.

**Section S9. The dependence of the bandgap *vs* the Zeeman field in 4 QLs and 7 QLs BST**

The computational details related to DFT/Perdew-Burke-Ernzerhof (PBE) calculations and bulk energy bands are reported in Section S8, Supporting Information. In 4 QLs BST, there is a small but finite gap due to the interaction between the top and the bottom SSs. In order to model the magnetic ordering in BST, one may consider a [001] Zeeman term that induces BC. Figure S9a,b show the numerically calculated band structures of 4 QLs BST with Zeeman fields of 30 meV and 50 meV. Our calculations show that the bandgap at the Γ point increases when a larger Zeeman splitting is used. Figure S9c shows the Zeeman-field dependence of the bandgap value at the Γ point. To realize the QAHE, the Zeeman field is required to exceed a quantum critical point ($q_c$); otherwise, 4 QLs BST are a normal insulator. States of BST exhibit a Chern number ($C$) of zero with the Zeeman field up to $q_c$, at which the bulk gap closes. Eventually, the gap re-opens for a Zeeman field larger than $q_c$, such that the topological phase transition occurs. In general, the nontrivial bandgap of BST increases considerably by increasing the strength of the Zeeman field. This gap makes this material suitable for studying the interplay between magnetism and band topology. Calculations on 7 QLs BST are shown in Figure S9d–f, and these results are similar to those of 4 QLs BST. The smaller $q_c$ value in Figure S9f may be attributed to the smaller hybridization gap in the 7 QLs BST. Therefore, the Zeeman field required to band inversion in 7 QLs BST is smaller than that in 4 QLs BST.

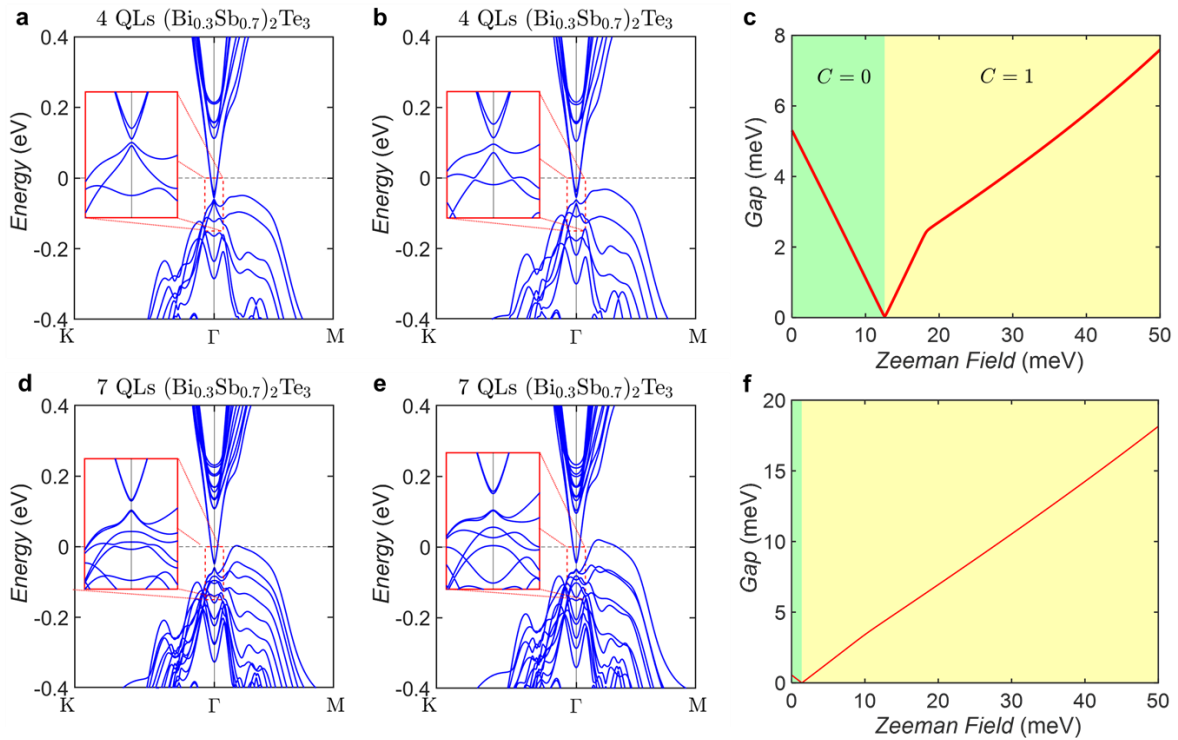

**Figure S9.** *Ab initio* band structure calculations along high symmetry lines of (a) 4 QLs BST with 30 meV Zeeman field, (b) 4 QLs BST with 50 meV Zeeman field, (d) 7 QLs BST with 30 meV Zeeman field, and (e) 7 QLs BST with 50 meV Zeeman field. Insets in (a), (b), (d), and (e) are the fine band structures around the $\Gamma$ point. (c) and (f) The dependence of the bandgap *vs* the Zeeman field in 4 QLs BST and 7 QLs BST, respectively. There is a clear phase transition for the topological feature, *i.e.*, $C = 0$ ($C = 1$) for the green (yellow) part.

**Section S10. Method for extracting $R_{AHE}$ and $R_{THE}$ values**

The Hall trace was the summation of OHE, AHE, and THE signals. After subtracting the linear OHE background, the remained signals are AHE and THE, as shown in Figure S10. The AHE signal can be described by $R_{AHE} = R_{AHE-max} \tanh(\frac{H \pm H_{c1}}{H_0})$, where $R_{AHE-max}$ is the $R_{AHE}$ value we plot in Figure 4c, $H_{c1}$ is the coercive field corresponding to the bottom SS, and $H_0$ is the fitting parameter.[7] Hence, the $R_{AHE}$ and $R_{THE}$ values of the loops are more precisely defined in *sample E*, where the AHE and THE coexist.

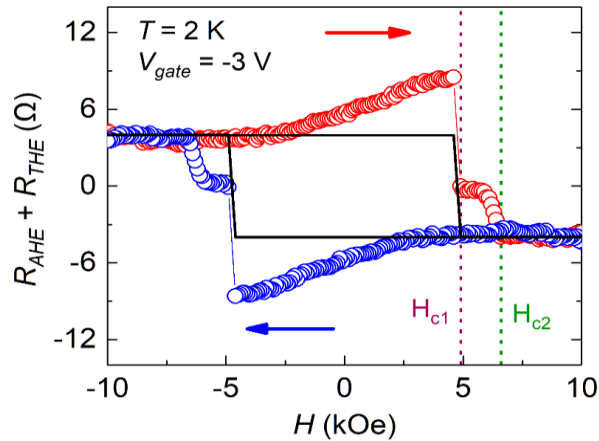

**Figure S10.** Coexistence of AHE and THE signals in *sample E* with $V_{gate}$ = -3 V at 2 K. The scattered points are the Hall data after subtracting a linear OHE background; the solid line is the fitted AHE component.

### Section S11. Discussion on the Absence of QAHE at 2 K

Recently, a Berry-curvature-driven anomalous Hall regime at temperatures of several Kelvin was reported in a MnBi$_2$Te$_4$/Bi$_2$Te$_3$ superlattice.[8] So far, QAHE by the approach of MPE was observed only in a ZCT/BST/ZCT sandwiched heterostructure below 0.1 K,[9] slightly higher than 30 mK using the magnetic doping approach by C.-Z. Chang *et al*.[10] Therefore, the measurement temperature of 2 K in our experiments could be one of the reasons that QAHE was not observed. Although the very large AHE was observed at room temperature, another challenge remained to observe QAHE in the BST/EuIG bilayer structure. The Zeeman field induced by the MPE decays rapidly from the bottom surface into the bulk of BST, indicating that the exchange interaction at the top surface of BST is extremely weak. Despite a large Zeeman field at the bottom interface, the Zeeman field at the top surface could still be weaker than the $q_c$, where the Chern number is zero. (Figure S9c) Hence, to realize QAHE *via* the MPE, the heterostructure must be a sandwiched heterostructure in the form of FI/TI/FI.[9,11] Now that we have elucidated the behavior of BST/EuIG bilayers, a new tri-layer heterostructure may be designed, in which both the top and bottom surfaces are magnetized to forms two independent SSs in TI to seek QAHE.

REFERENCES

1. Hikami, S.; Larkin, A. I.; Nagaoka, Y. Spin-Orbit Interaction and Magnetoresistance in the Two Dimensional Random System. *Prog. Theor. Phys.* **1980**, *63*, 707.

2. Zhang, H.; Liu, C.-X.; Qi, X.-L.; Dai, X.; Fang, Z.; Zhang, S.-C. Topological Insulators in $Bi_2Se_3$, $Bi_2Te_3$ and $Sb_2Te_3$ with a Single Dirac Cone on the Surface. *Nat. Phys.* **2009**, *5*, 438–442.

3. Benalcazar, W. A.; Bernevig, B. A.; Hughes, T. L. Quantized Electric Multipole Insulators. *Science* **2017**, *357*, 61–66.

4. Benalcazar, W. A.; Bernevig, B. A.; Hughes, T. L. Electric Multipole Moments, Topological Multipole Moment Pumping, and Chiral Hinge States in Crystalline Insulators. *Phys. Rev. B* **2017**, *96*, 245115.

5. Wieder, B. J.; Bernevig, B. A.; The Axion Insulator as a Pump of Fragile Topology. https://arxiv.org/abs/1810.02373 (accessed on 2018-10-04) arXiv:1810.02373v1.

6. Bryant, G. W. Surface States of Ternary Semiconductor Alloys: Effect of Alloy Fluctuations in One-Dimensional Models with Realistic Atoms. *Phys. Rev. B* **1985**, *31*, 5166.

7. Li, P.; Ding, J.; Zhang, S. S.-L.; Kally, J.; Pillsbury, T.; Heinonen, O. G.; Rimal, G.; Bi, C.; DeMann, A.; Field, S. B.; Wang, W.; Tang, J.; Jiang, J. S.; Hoffmann, A.; Samarth, N.; Wu, M. Topological Hall Effect in a Topological Insulator Interfaced with a Magnetic Insulator. *Nano Lett.* **2021**, *21*, 84–90.

8. Deng, H.; Chen, Z.; Wołoś, A.; Konczykowski, M.; Sobczak, K.; Sitnicka, J.; Fedorchenko, I. V.; Borysiuk, J.; Heider, T.; Pluciński, Ł.; Park, K.; Georgescu, A. B.; Cano, J.; Krusin-Elbaum, L. High-Temperature Quantum Anomalous Hall Regime in a $MnBi_2Te_4$/$Bi_2Te_3$ Superlattice. *Nat. Phys.* **2021**, *17*, 36–42.

9. Watanabe, R.; Yoshimi, R.; Kawamura, M.; Mogi, M.; Tsukazaki, A.; Yu, X. Z.; Nakajima, K.; Takahashi, K. S.; Kawasaki, M.; Tokura, Y. Quantum Anomalous Hall Effect Driven by Magnetic Proximity Coupling in All-Telluride Based Heterostructure. *Appl. Phys. Lett.* **2019**, *115*, 102403.

10. Chang, C.-Z.; Zhang, J.; Feng, X.; Shen, J.; Zhang, Z.; Guo, M.; Li, K.; Ou, Y.; Wei, P.; Wang, L.-L.; Ji, Z.-Q.; Feng, Y.; Ji, S.; Chen, X.; Jia, J.; Dai, X.; Fang, Z.; Zhang, S.-C.; He, K.; Wang, Y.; Lu, L.; Ma, X.-C.; Xue, Q.-K. Experimental Observation of the Quantum Anomalous Hall Effect in a Magnetic Topological Insulator. *Science* **2013**, *340*, 167–170.

11. Hou, Y.; Kim, J.; Wu, R. Magnetizing Topological Surface States of $Bi_2Se_3$ with a $CrI_3$ Monolayer. *Sci. Adv.* **2019**, *5*, eaaw1874.